\documentclass[12pt]{article} 
\usepackage{abstract}
\usepackage[utf8]{inputenc} 
\usepackage{amsxtra} 
\usepackage{amsthm}
\usepackage{amssymb}
\usepackage{amsfonts}
\usepackage{graphicx}
\usepackage{rotating}
\usepackage{color}
\usepackage{relsize}
\usepackage{epsfig}
\usepackage{subcaption} 
\usepackage{verbatim}
\usepackage[printonlyused]{acronym} 
\usepackage{setspace}
\usepackage{epstopdf}
\usepackage{authblk}
\usepackage{sectsty}
\usepackage[colorinlistoftodos]{todonotes}
\usepackage[colorlinks=true,allcolors=black]{hyperref}
\usepackage{textcomp,marvosym}
\usepackage{soul}
\usepackage{algorithm}
\usepackage{algorithmicx}
\usepackage{algpseudocode}
\usepackage{url}
\usetikzlibrary{shapes,arrows,automata,positioning}
\usepackage{mathtools}

\sectionfont{\fontsize{12}{15}\selectfont} 
\subsectionfont{\fontsize{10}{12}\selectfont}

\usepackage{geometry} 
\definecolor{blue}{RGB}{0, 0, 150}
\definecolor{green}{RGB}{0,150,0}
\definecolor{red}{RGB}{200, 0, 0}
\definecolor{black}{RGB}{0, 0, 0}
     
\renewcommand{\l}{\left}
\renewcommand{\r}{\right}

\newcommand{\dd}{\text{\,d}}
\definecolor{Purpleee}{RGB}{91,40,224}
\newcommand{\bfv}[1]{{\bf #1}}
\newcommand{\bfvL}{\bfv{L}}
\newcommand{\bfvC}{\bfv{C}}
\newcommand{\bfvK}{\bfv{K}}
\newcommand{\bfvM}{\bfv{M}}
\newcommand{\bfvF}{\bfv{F}}
\newcommand{\bfvR}{\bfv{R}}
\newcommand{\bfvW}{\bfv{W}}
\newcommand{\bfvV}{\bfv{V}}
\newcommand{\bfvT}{\bfv{T}}

\newcommand{\bfvO}{\bfv{\Omega}}
\newcommand{\bfvLamb}{\bfv{\Lambda}}
\newcommand{\bfvPhi}{\bfv{\Phi}}

\newcommand{\bfvg}{\bfv{g}}

\newcommand{\Hil}{\mathcal{H}}
\newcommand{\Hilg}{\mathcal{H}_\bfvg}
\newcommand{\Hilgbar}{\mathcal{H}_{\bar{\bfvg}}}
\newcommand{\MM}{{\mathcal{M}}}
\newcommand{\MMbar}{{\bar{\mathcal{M}}}}

\newtheorem{lemma}{Lemma}

\newtheorem{theorem}{Theorem}
\newtheorem{corollary}{Corollary}

\begin{document}

\title{Data-driven learning for the Mori--Zwanzig formalism: a generalization of the Koopman learning framework}
\author[1,*]{Yen Ting Lin}
\author[2]{Yifeng Tian}
\author[1]{Marian Anghel}
\author[2]{Daniel Livescu}
\affil[1]{Information Sciences Group, Computer, Computational and Statistical Sciences Division (CCS-3), Los Alamos National Laboratory, Los Alamos, NM 87545, USA}	
\affil[2]{Computational Physics and Methods Group, Computer, Computational and Statistical Sciences Division (CCS-2), Los Alamos National Laboratory, Los Alamos, NM 87545, USA}
\affil[*]{Corresponding author:  yentingl@lanl.gov}
\date{\today}

	\maketitle
    \begin{abstract}
{   A theoretical framework which unifies the conventional Mori--Zwanzig formalism and the approximate Koopman learning of deterministic dynamical systems from noiseless observation is presented. In this framework, the Mori--Zwanzig formalism, developed in statistical mechanics to tackle the hard problem of construction of reduced-order dynamics for high-dimensional dynamical systems, can be considered as a natural generalization of the Koopman description of the dynamical system. We next show that similar to the approximate Koopman learning methods, data-driven methods can be developed for the Mori--Zwanzig formalism with Mori's linear projection operator. We have developed two algorithms to extract the key operators, the Markov and the memory kernel, using time series of a reduced set of observables in a dynamical system. We have adopted the Lorenz `96 system as a test problem and solved for the above operators. These operators exhibit complex behaviors, which are unlikely to be captured by traditional modeling approaches in Mori--Zwanzig analysis. The nontrivial Generalized Fluctuation Dissipation relationship, which relates the memory kernel with the two-time correlation statistics of the orthogonal dynamics, was numerically verified as a validation of the solved operators. We present numerical evidence that the Generalized Langevin Equation, a key construct in the Mori--Zwanzig formalism, is more advantageous in predicting the evolution of the reduced set of observables than the conventional approximate Koopman operators. \vspace{6pt} \vspace{6pt}}
	\end{abstract}
\vspace{10pt}

\section{Introduction}

The Mori--Zwanzig formalism \cite{mori1965transport,zwanzig1973nonlinear,zwanzig2001nonequilibrium,evans2007statistical} was first developed in statistical physics for the difficult task of constructing coarse-grained models from high-dimensional microscopic models. The goal of model coarse-graining is to construct  equations describing the evolution of a smaller set of variables which are measurable, or quantities of our interests. These quantities are often referred to as the relevant or resolved variables. For example, a microscopic model can be all-atom molecular dynamics simulation of a protein in a solvent, and a relevant variable can be the distance between two atoms of our interests. It is desirable to construct a closed dynamical system which include only the resolved variables without the information of other degrees of freedom. On the one hand, it is easier to perform analysis on a closed lower-dimensional system and to shed 
important insights on the interactions between the resolved variables. On the other hand, it is more efficient to simulate the reduced-dimensional system computationally. 

The major challenge of coarse-graining modeling is that the resolved variables may be influenced by the unresolved variables. In the above example, the distance between the two specific atoms may be influenced by nearby water molecules that are not in the set of resolved variables. To solve this difficult closure problem, Mori \cite{mori1965transport} and Zwanzig \cite{zwanzig1973nonlinear} developed the projection-based methods to express the effect of the unresolved variables in terms of the resolved ones. The Generalized Langevin Equation, the main result of the Mori--Zwanzig formalism, decomposes the evolutionary equations of the resolved variables into three parts: a Markov term, which captures the interaction within the resolved variables, a memory term, which is history-dependent and captures the interactions between the resolved and the unresolved variables, and a term representing the orthogonal dynamics, which captures the unknown initial condition of the unresolved variables. 
Although the Generalized Langevin equation is formally exact, it is challenging to theoretically derive closed-form expressions of these terms in the Generalized Langevin Equation without approximations. 
Conventionally, the applications of the Mori--Zwanzig formalism rely on modeling self-consistent operators based on the mathematical structure of the Generalized Langevin Equation  \cite{stinis2004stochastic,horenko07data,chen14computation,li2015incorporation,lei2016data,li2017computing,parish2017non,wang2019implicit,hudson20coarse}.

In a seemingly unrelated research area, approximate Koopman learning methods such as Dynamic Mode Decomposition \cite{schmid2010DMD,schmid2011applications} and Extended Dynamic Mode Decomposition \cite{williams2015data}, have been actively developed for data-driven modeling of dynamical systems. The general idea is that by collecting enough data of a dynamical system, possibly from a high-fidelity simulation of the microscopic system, one would be able to learn important features in the dynamics, e.g., spectral \cite{Mezic2005a} and dynamic modes \cite{schmid2010DMD,schmid2011applications,williams2015data}. The theoretical foundation of these methods, the Koopman theory, is a formulation for general dynamical systems \cite{Koopman255,Koopman315}. Instead of the typical description of a possibly nonlinear system in the physical space, in Koopman theory, the dynamics are described as a linear dynamical system of the functions of the observables in an infinite-dimensional Hilbert space. Because the interactions in this framework are linear (but with a caveat that the space is infinite dimensional), learning from data is a convex problem which is easier to solve, in contrast to the nonlinear regression in the conventional physical-space picture. 

The major aim of this article is to bridge these two seemingly disconnected research areas in dynamical systems: the Mori--Zwanzig formalism and the approximate Koopman learning. We will establish that \emph{the Mori--Zwanzig formalism with Mori's linear projector is functionally identical to the approximate Koopman learning methods in a shared Hilbert space}. This connection helps to bring the advantage of one research area to another. On the one hand, the approximate Koopman learning methods can be generalized for data-driven modeling of the operators in Mori--Zwanzig formalism. As will be seen in this article, the operators describing the Markov and memory terms in the Generalized Langevin Equation can be numerically learned from the simulation of the microscopic system. Surprisingly, these operators, inferred from data, are highly nontrivial and are unlikely to be modeled accurately without in-depth knowledge of the system. On the other hand, the memory terms in the Mori--Zwanzig formalism can be considered as higher-order corrections of the Koopman learning methods. We will show that by including the memory kernel and the history of the resolved variables, the Generalized Langevin Equation predicts more accurately than the Extended Dynamic Mode Decomposition. 

This article is organized by the following structure. In Sec.~\ref{sec:preliminaries}, we provide a gentle introduction to the two descriptions of a dynamical system: the description of a finite-dimensional, but possibly nonlinear dynamics in the physical space, and the Koopman description of an infinite-dimensional, but linear dynamics of the observables. For completeness, we include a self-contained introduction and review of the Mori--Zwanzig formalism in Sec.~\ref{sec:MZ}. We establish that the Mori--Zwanzig formalism is a generalization, in the sense that it contains the higher-order memory effect, of the Extended Dynamic Mode Decomposition (EDMD, \cite{williams2015data}) in Sec.~\ref{sec:koopman}. Two novel algorithms, motivated by the EDMD to extract the key operators in the Mori--Zwanzig formalism by simulation data, are presented in Sec.~\ref{sec:algos}. We perform numerical experiments on a Lorenz `96 model \cite{lorenz1996predictability} and present the results in Sec.~\ref{sec:experiments}. In Sec.~\ref{sec:errorAnalysis}, we demonstrate the advantage of the Mori--Zwanzig formalism over the conventional EDMD in predicting dynamical systems into the future. Finally, we provide a discussion and future outlook in Sec.~\ref{sec:discussion}.

\section{Preliminaries}\label{sec:preliminaries}

There exist two equivalent formulations to describe a dynamical system. In the first formulation \cite{abraham1978foundations,strogatz2000nonlinear}, the system is characterized by a collection of \emph{physical-space variables}, often termed as the state of the system. For example, a physical-space variable can be one component of the position of an atom in a many-particle system, or one component of the velocity field at a specific location in a fluid dynamical system. The aim of this first formulation is to describe the evolution of these physical-space variables. Suppose the state of the system is fully characterized by $N$ physical-space variables $\phi_i$, $i=1,\ldots,N$, and we denote the state of the system at time $t$ by $\bfvPhi\l(t\r):=\l[\phi_1\l(t\r), \ldots, \phi_N\l(t\r)\r]^T$, an $N\times 1$ column vector. Then, the evolution of the variables in the physical space (assumed to be $\mathbb{R}^N$ for simplicity) is described by the deterministic evolutionary equations
\begin{equation}
\frac{\dd }{\dd t} \phi_i\l(t\r)= R_i \l(\bfvPhi\l(t\r), t \r),\quad \text{and} \quad \bfvPhi\l(0\r)=\bfvPhi_0, \quad i=1, \ldots N, \label{eq:phiEvo}
\end{equation}
where the flow $R_i:\mathbb{R}^N\rightarrow \mathbb{R}$ is a function which maps the state $\bfvPhi$ to a real number that characterizes the velocity of the physical-space variable $\phi_i$ at time $t$, and $\bfvPhi_0$ is the given $N\times 1$ column vector specifying the initial condition of the system's state. In this article, we exclusively consider autonomous dynamical systems, where $R_i$ does not explicitly depends on the time $t$. Thus, the evolutionary equation \eqref{eq:phiEvo} can be written in a terse form $\dot{\bfvPhi} = \bfvR\l(\bfvPhi\r)$, where the flow $\bfvR$ is defined as $\l[R_1\l(\bfvPhi\l(t\r)\r), \ldots, R_N\l(\bfvPhi\l(t\r)\r)\r]^T$ and is implicitly time-dependent. In general, the flow $\bfvR$ can be nonlinear in $\bfvPhi$. 

In the second formulation proposed by Koopman \cite{Koopman255,Koopman315}, the system is characterized by a collection of \emph{observables} which are functions of the physical-space variables. For example, an observable can be a component of the total angular momentum of a subset of all atoms in a particle system, or the locally averaged density in a fluid dynamical system. The Koopmanian formulation describes how observables evolve in an infinite-dimensional Hilbert space $\mathcal{H}$, which is composed of all the possible observables. The advantage of this formulation is that the evolution of the observables, which is a vector in the infinite dimensional Hilbert space $\mathcal{H}$, is \emph{always linear}, even for systems that are nonlinear in the physical-space picture. The disadvantage of this formulation is that the state space of the system, which consists of all possible observables, is infinite dimensional.

To illustrate the difference of the formulations, we consider a one-dimensional nonlinear dynamical system in the physical-space formulation: $\dot{\phi}(t) = R\l(\phi\l(t\r)\r)$ and $\phi\l(0\r):=\phi_0$, where $R(x):=-x^2$ is a nonlinear function and $\phi_0$ is the initial condition. While the analytic solution exists for this simple problem ($\phi\l(t\r) = 1/(t+1/\phi_0)$), it is challenging to derive the closed-form solution for general multidimensional ($N>1$) nonlinear dynamical systems. Note that $\phi(t)$ is a nonlinear function of the initial condition $\phi_0$. In the Koopman formulation, the dynamics are characterized by observables of $\phi$. It is sufficient for us to consider a set of observables which will serve as the basis functions. For this example, we consider $g_{k}(\phi) :=\phi^{k}$, $k\in \mathbb{Z}_+$. Other observables can be expressed as a weighted linear superposition of these basis functions via Taylor series expansion. In contrast to the first formulation, Koopman's theory describes the dynamics of the basis functions $g_k$:
\begin{equation}
\frac{\dd}{\dd t} g_k\l(t\r):= \frac{\dd}{\dd t} \l[g_k \circ \phi\l(t\r)\r] = \frac{\dd g_k}{\dd \phi} \cdot \frac{\dd\phi}{\dd t} = k \phi^{k-1}\l(t\r)\cdot\l[-\phi^2\l(t\r)\r] = -k g_{k+1}\l(t\r). 
\end{equation}
Throughout this article, we will use the symbol $\circ$ to denote the composite functions. Here, the observable functions $g_k$ are functions of the physical-space variable $\phi$, which is a function of the physical time $t$. The dynamics of $g_k(t)$ is always linearly dependent to $g_{k+1}(t)$, but is not closed unless the system involves infinitely many $k$'s. The evolution of lower-order nonlinearity involves higher-order nonlinearity, similar to the common phenomenon in Carleman linearization \cite{carleman1932,kowalski91nonlinear} and moment expansion methods \cite{ale2013general,Schnoerr2015comparison}. Nevertheless, we can choose two simple functions $g_0(\phi):=\phi^0$ and $g_{-1}\l(\phi\r)=\phi^{-1}$ and their dynamics are closed:
\begin{equation}
\frac{\dd }{\dd t} \l[\begin{array}{c} g_0 \l(t\r) \\ g_{-1}\l(t\r) \end{array}\r] = \l[\begin{array}{cc} 0 & 0 \\ 1 & 0 \end{array}\r]\l[\begin{array}{c} g_0 \l(t\r) \\ g_{-1}\l(t\r) \end{array}\r], \quad \text{and} \quad \l[\begin{array}{c} g_0 \l(0\r) \\ g_{-1}\l(0\r) \end{array}\r] = \l[\begin{array}{c} 1 \\ \phi_0^{-1} \end{array}\r].
\end{equation}
The above \emph{linear} ordinary differential equations are solved to derive the analytical solution of $g_{-1}\l(\phi\l(t\r)\r)=t+\phi_0^{-1}$, which is then used to calculate $\phi\l(t\r)$. The choice of this invariant set of functions is not unique. We can even choose a smaller set which contains only one observable $g_e  (\phi):=\exp(-1/\phi)$, which satisfies a one-dimensional linear ordinary equation $\dot{g_e}(t) = - g_e(t)$. 

The above examples illuminate two key features of the Koopman theory. First, the dynamics of observables are always linearly dependent on other observables. Secondly, to derive closed-form solution in the Koopman theory is equivalent to identifying a set of observables whose dynamics are invariant in a subspace which is linearly spanned by the set of the observables. In general, it is challenging to identify the finite set of observables that closes the dynamics, and one has to resort to approximation methods to close the system. In the next section, we illustrate how the Mori--Zwanzig formalism leverage the projection operators to close the dynamics.

\section{The Mori--Zwanzig Formalism} \label{sec:MZ}

Here, we provide a review to the Mori--Zwanzig formalism. For completeness, we provide two comprehensive derivations of the major result of the Mori--Zwanzig formalism, the Generalized Langevin Equation (GLE). We begin with the operator algebraic derivation \cite{chorin00optimal,AlexandreJ.Chorin2002,falkena2019DerivationDelayEquation} in Sec.~\ref{sec:operatorAlgebra}. To make the connection to the Koopman representation of the dynamics, we provide an alternative derivation of the GLE based on the the Koopman eigenfunctions in Sec.~\ref{sec:intuitiveDerivation}. Although the first derivation is terse and elegant, it is not easy to build intuition to understand the action of operators in the GLE. The second derivation has two advantages: (1) it provides a more transparent representation of the Mori--Zwanzig operators and (3) its terminology naturally bridges to approximate Koopman analysis such as EDMD \cite{williams2015data}. In fact, the second approach's geometric representation in the functional space is identical to Mori's original construct \cite{mori1965transport}, and the derivation is very close to the variation of constant method presented in Zwanzig's own derivation \cite{zwanzig2001nonequilibrium}. We will thus adopt the terminology of the second derivation throughout the rest of the paper. After the GLE is set up, we provide a geometric interpretation of the GLE in Sec.~\ref{sec:geometry} and a detailed discussion on the projection operator in Sec.~\ref{sec:projectionOperator}. In Sec.~\ref{sec:CE+OP} we discuss the consequence of the GLE on the evolutionary equations of the covariance matrices and the projected image. Sec.~\ref{sec:GFD} is dedicated to the emergence of the self-consistent generalized fluctuation dissipation relationship. We conclude the review to the Mori--Zwanzig by remarking its applicability to discrete-time dynamics in Sec.~\ref{sec:dMZ}.  
 
\subsection{Operator algebraic derivation of the Generalized Langevin Equation} \label{sec:operatorAlgebra}
We begin with the evolutionary equation \eqref{eq:phiEvo}, where the flow field $\mathbb{R}:\mathbb{R}^N\rightarrow \mathbb{R}^N$ is assumed to be locally Lipschitz continuous such that a unique $\bfvPhi(t)$ exists $\forall t\ge 0$. More generally, the state space can be any compact Riemannian manifold endowed with the Borel $\sigma$-algebra and a measure; for brevity, we will consider the state space as $\mathbb{R}^N$ below. The solution $\bfvPhi(t;\bfvPhi_0)$ nonlinearly maps the initial condition $\bfvPhi_0$ to the phase-space configuration at physical time $t$. Next, one defines the Liouville operator $\mathcal{L}:=\sum_{i=1}^N R_i\l(\bfv{x}\r) \partial_{x_i} $, with a dummy variable $\bfv{x}\in\mathbb{R}^N$, and considers the following partial differential equation (PDE)
\begin{subequations} \label{eq:PDE}
\begin{align}
    \frac{\partial \psi(t, \bfv{x})}{\partial t} ={}& \mathcal{L} \psi(t, \bfv{x}), \\
    \psi(0,\bfv{x}) ={}& g\l(\bfv{x}\r),
\end{align}
\end{subequations}
where $g:\mathbb{R}^N \rightarrow \mathbb{R}$ is real-valued function of the state of the system, $\bfv{x}\in \mathbb{R}^N$. One can show that, the solution to the above first-order PDE is $u(t, \bfv{x}) \equiv g\l(\bfvPhi\l(t; \bfv{x}\r)\r)$ by the method of characteristics. Note that $\bfv{x}\in\mathbb{R}^N$ can be any initial state. Thus, solving the above linear PDE fully solves the function $g$ evaluated at the trajectory of the nonlinear system given any initial condition $\bfvPhi_0$: $g\l(\bfvPhi\l(t; \bfvPhi_0\r)\r)=\psi(t, \bfvPhi_0)$. We adopt the slightly abused notation in published literature \cite{chorin00optimal,AlexandreJ.Chorin2002,falkena2019DerivationDelayEquation} and denote the solution $\psi\l(t,\bfv{x}\r)$ with the initial condition $\psi(0, \bfv{x}) = g(\bfv{x})$ by $g(t, \bfv{x})$. A special choice of $g$ is $g(\bfv{x}):=x_i$, that $g$ extracts the $i^\text{th}$ component of the multivariate vector. In this case $\psi(\bfvPhi_0,t)$ is the solution of the $i^\text{th}$ component, $\psi(t,\bfvPhi_0)\equiv \phi_i \l(t;\bfvPhi_0\r)$. The semigroup notation tersely represents the solution of the above PDE \eqref{eq:PDE} as $\psi(t,\bfv{x}) = e^{t \mathcal{L}} g(\bfv{x})$, with an evolutionary equation 
\begin{equation}
    \frac{\partial}{\partial t} \l[e^{t \mathcal{L}} g \r](\bfv{x}) =\l(\mathcal{L} e^{t\mathcal{L}}  g \r)(\bfv{x})= \l(e^{t\mathcal{L}} \mathcal{L} g \r)(\bfv{x}). \label{eq:semigroup}
\end{equation}
The above equation applies to any $\bfv{x}\in \mathbb{R}$ so ``$\l(\bfv{x}\r)$'' is often neglected in calculations. 

Conventionally, the goal of Mori--Zwanzig procedure is to construct the evolutionary equations for a set of components $\hat{\bfv{\phi}}:=\l\{\phi_i\r\}_{i=1}^M$, $M<N$, referred to as the resolved components. These are the components which we can measure as the dynamics move forward in time. Because the state space $\mathbb{R}^N$ can be fully characterized by $N$ coordinates $\phi_i$, $i=1\ldots N$, knowing $M<N$ resolved observables could not fully specify the system's state for constructing a closed dynamical system. Consequently, one would need to postulate another $N-M$ under-resolved components, often denoted by $\tilde{\phi}$. Mori--Zwanzig procedure proceeds with a postulated joint distribution $\dd \mu \equiv \rho(\bfv{x}) \dd^N \bfv{x}$, where $\rho$ is the probability density and $\dd^N \bfv{x}$ is the Borel measure in $\mathbb{R}^N$, for asserting the initial distribution of the under-resolved $\tilde{\phi}$ conditioned on a given set $\hat{\phi}$. The choice of $\dd \mu$ is model-specific but often the equilibrium (or non-equilibrium stationary) distribution of the system. Despite the conventional choice of using the components of the state (i.e., $g(\bfv{x})=x_i$) as the resolved and under-resolved observables, $g$ can be any function of the state (for example, $g$ can be the center of mass of a molecule whose full configurations are specified by the position and momentum of all its atoms.) A technical condition on $g$ is that it has to be $L^2$-integrable with respect to the measure $\dd \mu$ for constructing an inner product of a Hilbert space in which Mori--Zwanzig formalism operates. 
 
Mori--Zwanzig procedure proceeds with a specified projection operator $\mathcal{P}$, which maps a function of the full-space configuration, $g:\mathbb{R}^N \rightarrow \mathbb{R}$, to a function of only the resolved observables $\mathcal{P} g:\mathbb{R}^M \rightarrow \mathbb{R}$, assumed to be $L^2$-integrable with respect to $\dd \mu$. The complement of the projection operator is defined as $\mathcal{Q}:=I-\mathcal{P}$. Applying the Dyson identity \cite{evans2008StatisticalMechanicsNonequilibrium}
\begin{equation}
    e^{t(A+B)} = e^{tB} + \int_0^t e^{(t-s)(A+B)}A e^{sB} \dd s
\end{equation}
to operators $A:=\mathcal{PL}$ and $B:=\mathcal{QL}$, one obtains
\begin{equation}
    e^{t\mathcal{L}} = e^{t\mathcal{QL}} + \int_0^t e^{(t-s)\mathcal{L}}\mathcal{PL} e^{s\mathcal{QL}} \dd s.
\end{equation}
The operator is applied to Eq.~\ref{eq:semigroup}, resulting in the following expression for any $g$ with the fact that $\mathcal{P}+\mathcal{Q}=I$:
\begin{align}
    \frac{\dd}{\dd t}\l[e^{t\mathcal{L}} g\r] ={}& \l[e^{t\mathcal{L}} \mathcal{L} g\r]=  \l[e^{t\mathcal{L}} \l(\mathcal{P}+\mathcal{Q}\r) \mathcal{L} g\r]  \\
        = {}& \l[e^{t\mathcal{L}} \mathcal{PL} g\r] + \l[e^{t\mathcal{QL}} \mathcal{QL} g\r]+ \int_0^t e^{(t-s)\mathcal{L}} \l[\mathcal{PL} e^{s\mathcal{QL}} \mathcal{QL} g\r]  \dd s \nonumber
\end{align}
Specifically for $g\l(\bfv{x}\r)=x_i$, $i=1\ldots M$, one define the Markov transition $M_i(\hat{\bfv{x}}):=\l[\mathcal{P} R_i \r]\l(\bfvPhi\l(t,\bfv{x}\r)\r)$, the orthogonal dynamics $F_i(t, \bfv{x}):=\l[e^{t\mathcal{QL}}\mathcal{QL}g\r](\bfv{x})$ and the memory function $K_i(t,\bfv{\hat{x}}):= - \l[\mathcal{PL} F_i \r]\l(t,\bfv{x}\r)$ to obtain the the generalized Langevin equation (GLE) describing the evolution of resolved components given an initial condition $\bfvPhi_0$:
\begin{equation}
    \frac{\dd}{\dd t} \hat{\phi}_i(t, \bfvPhi_0)
    =M_i\l(\hat{\bfvPhi}\l(t,\bfvPhi_0\r)\r) - \int_0^t K_i\l(\hat{\bfvPhi}\l(t, \bfvPhi_0\r), t-s\r) \dd s + F_i(t, \bfvPhi_0). 
\end{equation}
Note that we follow the original sign convention that Mori \cite{mori1965transport} and Zwanzig \cite{zwanzig1973nonlinear} adopted: the memory term is with a negative sign, contrast to later publications \cite{chorin00optimal,AlexandreJ.Chorin2002,falkena2019DerivationDelayEquation} in which the memory term was defined with a positive sign. 

The difference between Mori and Zwanzig is their choice of the projection operator. With Mori's construction \cite{mori1965transport}, one relies on an inner product defined as 
\begin{equation}
    \l\langle f, g \r\rangle = \int_{\mathbb{R}^N} f\l(\bfv{x}\r) g\l(\bfv{x}\r) \rho\l(\bfv{x}\r) \dd^N \bfv{x}, \quad f, g\in L^2\l(\mu \r) \label{eq:MoriInner}
\end{equation}
to define a projection operator given a set of resolved observables $\hat{\phi}=\l\{\phi_i \r\}_{i=1}^M$:
\begin{equation}
    \l[\mathcal{P} f\r] \l(\hat{\phi} \r) := \sum_{i,j=1}^M \l\langle f, \phi_i \r\rangle \l[\bfvC^{-1}_0\r]_{i,j} \phi_j.
\end{equation}
where $\bfvC^{-1}(0)$ is the inverse of an $M\times M$ matrix $\bfv{C}_0$ whose $(i,j)$ entry is $\l\langle \phi_i, \phi_j \r\rangle$. A more geometric interpretation of Mori's projector will be presented in Sec.~\ref{sec:projectionOperator}. Note that $\mathcal{P} f $ is a linear function of the resolved observables, $\phi_j$, $j=1\ldots M$, and thus Mori's projector is often referred to as a linear projection. In contrast, with the same set of observables, Zwanzig \cite{zwanzig1973nonlinear} does not rely on the inner product but relies on the direct marginalization of the under-resolved obeservables:
\begin{equation}
    \l[\mathcal{P} f\r] \l(\hat{\bfv{x}} \r) := \frac{\int_{\mathbb{R}^{N-M}} f\l(\hat{\bfv{x}}, \tilde{\bfv{x}}\r) \rho\l(\hat{\bfv{x}}, \tilde{\bfv{x}}\r) \dd \tilde{\bfv{x}} }{\int_{\mathbb{R}^{N-M}} \rho\l(\hat{\bfv{x}}, \tilde{\bfv{x}}\r) \dd \tilde{\bfv{x}}}.
\end{equation}
Note that the resulting function $\mathcal{P}f $ is generally nonlinear in the resolved observables, and thus, Zwanzig's projection is often referred to as the ``nonlinear projection''. Also termed as the nonlinear projection \cite{AlexandreJ.Chorin2002} and infinite-rank projection \cite{falkena2019DerivationDelayEquation}, Zwanzig's projection opeator can lead to a nonlinear Markov transition and nonlinear memory kernel in the Generalized Langevin Equation \cite{chorin00optimal,AlexandreJ.Chorin2002,hudson20coarse}.

After applying Mori's projection operator to $M$, $F$, and $K$, one obtains a linear GLE \cite{zwanzig2001nonequilibrium}: 
\begin{equation}
    \frac{\dd}{\dd t} \hat{\phi}_i(t, \bfvPhi_0)
    = \sum_{j=1}^M \l[\bfvM\r]_{i,j} \hat{\phi}_j\l(t, \bfvPhi_0\r) - \int_0^t \l[\bfvK(t-s)\r]_{i,j} \hat{\phi}_j\l(t, \bfvPhi_0\r) \dd s + F_i(t, \bfvPhi_0).  \label{eq:opGLE}
\end{equation} 
where $\bfvM$ is an $M\times M$ constant matrix, as well as $\bfvK(t)$, $t\ge 0$. We will illuminate the physical meaning of this linear GLE in Sec.~\ref{sec:geometry}. In this manuscript, we will focus on Mori's projection and its connection to the approximate Koopman learning methods.

\subsection{Generalized Langevin Equation in Koopman representation} \label{sec:intuitiveDerivation}

It is not easy to build intuitions from the terse derivation presented in the previous section \ref{sec:operatorAlgebra}. R.~Zwanzig even commented ``\emph{The derivation to be given here is based on abstract operator manipulations that were designed to get to the desired result as quickly as possible}'' and provided a more lengthy motivating derivation based on variation of constant method prior to the formal operator algebraic derivation. In this section, we aim to provide a similar derivation using Koopman representation of the dynamics for better understanding the GLE. By introducing a Koopman representation of the dynamics, we also aim to make a natural and formal connection between Mori--Zwanzig and Koopman formulations.

As shown in the motivating example in Sec.~\ref{sec:preliminaries}, in Koopman representation of the dynamics, one aims to describe the evolution of the observables, which are functions of the system's state $\bfvPhi$. In the space of all $L^2$-integrable observables, the evolution is always linear in other observables, but the dimensionality of the operating space can be infinite. Formally, given a measure $\dd \mu$, we denote the space of all $L^2$-integrable real-valued (can be generalized to complex-valued) observables of the state space by $\mathcal{F}=L^2\l(\mathbb{R}^N, \mu \r)$. Together with a defined inner product, such as \eqref{eq:MoriInner}, these functions form a Hilbert functional space $\Hil$, in which the functions evolve forward in time. The continuous-time Koopman operator $\mathcal{K}_t:\mathcal{F}\rightarrow \mathcal{F}$ is defined by
\begin{equation}
    \l(\mathcal{K}_t g \r) \l(\bfvPhi_0\r) = g \circ \bfvPhi \l(t; \bfvPhi_0\r) \equiv g \l(\bfvPhi \l(t; \bfvPhi_0\r)\r), \, \forall g\in \mathcal{F}, \,\forall \bfvPhi_0 \in \mathbb{R}^N. 
\end{equation}
In other words, Koopman operator $\mathcal{K}_t$ transforms the function $g$ to a function $\mathcal{K}_t g$ of any initial condition $\bfvPhi_0$. At any time $t\ge 0$, $\mathcal{K}_t g$ is equivalent to the observable $g$ evaluated at the solution of the dynamics $\bfvPhi\l(t; \bfvPhi_0\r)$, which is also a function of $\bfvPhi_0$. The above equation shows the \emph{dual representations} of the dynamics: The left hand side of the equation is the Koopman picture, analogous to the Heisenberg picture in quantum mechanics, that the observable is evolving (transformed by $\mathcal{K}_t$) forward in time and is always evaluated at the fixed state $\bfvPhi_0$, while the right hand side of the equation is the Perron--Frobenius picture, analogous to the Schr\"{o}dinger's picture in quantum mechanics, that the state is evolving forward in time ($\bfvPhi\l(t; \bfvPhi_0\r)$) and evaluated by a fixed observable $g$.

The linear Koopman operator $\mathcal{K}_t$ can be characterized by its eigenvalues and eigenfunction. A function $\phi:\mathbb{R}^N\rightarrow \mathbb{R}$ (or $\mathbb{C}$) is defined as a Koopman eigenfunction if it satisfies $\l(\mathcal{K}_t  \phi\r)= e^{\lambda t} \phi$. Here, we drop the ``as a function of initial-condition'' annotation ``$\l(\bfvPhi_0\r)$'' again. The space of the eigenfunctions are infinite-dimensional: given two pairs $(\lambda_1, \phi_1)$ and $(\lambda_2, \phi_2)$, one can generate infinitely many eigenfunctions $(m\lambda_1+n\lambda_2, \phi_1^m \phi_2^n)$, $m,n \in \mathbb{N}$. The infinitesimal generator of $\mathcal{K}_t$, $\lim_{t\downarrow 0}\l(\mathcal{K}_t-I\r)/t$ is the Liouville operator $\mathcal{L}:=\sum_{i=1}^N R_i \l(\bfv{x}\r) \partial_{x_i}$, which is the Lie derivative with respect to the flow field $\bfv{R}:\mathbb{R}^N \rightarrow \mathbb{R}^N$ \cite{mauroy2016GlobalStabilityAnalysisa}. Note that $\mathcal{L} \phi = \lambda \phi$. As such, the Koopman eigenfunctions are the eigenfunctions of the Liouville operator and are special initial data following a coherent evolution $\psi(t, \bfv{x})=\phi(\bfv{x}) \exp\l(\lambda t\r)$ by Eq.~\eqref{eq:PDE}. 

One aims to construct the evolution of a set of linearly independent observables, $\MM:=\l\{g_i \r\}_{i=1}^M$. Given a time $t\ge 0$, we would like to know how $g_i(t):=\mathcal{K}_t g_i$, a function of the initial condition $\bfvPhi_0$ parametrized by time $t$, changes with respect to time $t$. We remark that such a dynamical variable notation (``$g_i(t)$'') was first introduced by Mori \cite{mori1965transport} and has been the mainstream notation in the physics literature \cite{zwanzig2001nonequilibrium}. Using the modern Koopman notation, $g(t)$ is expressed as $\mathcal{K}_t g_i$, and with the algebraic notation in Sec.~\ref{sec:operatorAlgebra} as $g_i(t, \cdot)$. Note that $g_i(0)=\mathcal{K}_0 g_i=g_i$. Although the Koopman operator may contain a continuous spectrum \cite{Koopman255,neumann1932ZurOperatorenmethodeKlassischen,Mezic2005a} for chaotic dynamical systems, in this derivation, we consider systems with only point spectra for brevity. For these systems, the observables $g_i$ can be expressed as a linear combination of the countably infinite eigenfunctions \cite{rowley2009SpectralAnalysisNonlinear,williams2015data}:
\begin{equation}
    g_i = \sum_{j=1}^\infty v_{i,j} \phi_j, \, i=1\ldots M. \label{eq:KoopmanEigDecomposition}
\end{equation}
In contrast to the above equation which decompose a function into Koopman eigenfunctions, Mori--Zwanzig formalism utilizes the inner product in the Hilbert space to decompose the space into the subspace linearly spanned by the set of observables, $\Hilg:=\text{Span}(\MM)$, and an orthogonal subspace $\Hilgbar=\l\{\bar{g} \in \mathcal{F}: \l\langle \bar{g} , g_i\r\rangle=0, g_i \in \MM \r\}$. One proceeds with constructing a complete set of basis functions in $\Hil$, with a natural choice of using $\MM$ as the set of basis functions in $\Hilg$. One can then use the Gram-Schimidt process to construct the basis functions in the orthogonal space from the Koopman eigenfunctions, $\l\{\phi_i\r\}_{i=1}^{\infty}$. We denote this infinite set of basis functions by $\MMbar:=\l\{\bar{g}_i\r\}_{i=1}^\infty$. 
Similar to Eq.~\eqref{eq:KoopmanEigDecomposition}, we can decompose $\bar{g}_i$'s in terms of the eigenfunctions: $\bar{g}_i = \sum_{j=1}^\infty \bar{v}_{i,j} \phi_j, \, i\in \mathbb{N}.$
By construction, $\l\langle g_i, \bar{g}_j \r\rangle = 0$, $i \in \l\{1, \ldots, M\r\}$, $j \in \mathbb{N}$. Note that we did not require orthogonality between the basis functions in the same subspace, that is, $\l\langle g_i, g_j \r\rangle$ and $\l\langle \bar{g}_i, \bar{g}_j \r\rangle$ are not required to be $0$ if $i\ne j$. However, we assume linear independence between any of the pairs of the basis functions, and the combined set $\MM \cup \MMbar$ forms a complete set of basis functions in $\Hil$. Consequently, one can express any Koopman eigenfunction $\phi_i$ in terms of these new basis functions
\begin{equation}
    \phi_i = \sum_{j=1}^M \omega_{i,j} g_j + \sum_{j=1}^\infty \bar{\omega}_{i,j} \bar{g}_j. 
\end{equation}
Applying the Koopman operator $\mathcal{K}_t$ to the above equation, we obtain the relationship $\forall t\ge 0$, $\phi_i(t) = \sum_{j=1}^M \omega_{i,j} g_j(t) + \sum_{j=1}^\infty \bar{\omega}_{i,j} \bar{g}_j (t)$. Now, we can express the evolution of the basis functions $g_i$, $i=1\ldots M$:
\begin{align}
    \frac{\dd}{\dd t} g_i(t) = {}&\lim_{s\downarrow 0} \frac{\mathcal{K}_s g_i(t) - g_i(t) }{s} = \sum_{j=1}^\infty v_{i,j} \lambda_j e^{\lambda_j t} \phi_j = \sum_{i=0}^\infty v_{i,j} \lambda_j \phi_j(t)  \label{eq:giEvol} \\
    ={}& \sum_{\ell=1}^M \l(\sum_{j=1}^\infty  v_{i,j} \lambda_j \omega_{j,\ell} \r) g_\ell(t) + \sum_{\ell=1}^\infty \l(\sum_{j=1}^\infty v_{i,j} \lambda_j \bar{\omega}_{j,\ell}\r) \bar{g}_\ell(t), \nonumber  
\end{align}
and similarly for $\bar{g}_i$, $i\in \mathbb{N}$:
\begin{align}
    \frac{\dd}{\dd t} \bar{g}_i(t) = \sum_{\ell=1}^M \l(\sum_{j=1}^\infty  \bar{v}_{i,j} \lambda_j \omega_{j,\ell} \r) g_\ell(t) + \sum_{\ell=1}^\infty \l(\sum_{j=1}^\infty \bar{v}_{i,j} \lambda_j \bar{\omega}_{j,\ell}\r) \bar{g}_\ell(t).  \label{eq:giBarEvol}
\end{align}
We have established that at any time, the evolution of $g_i(t)$ and $\bar{g}_i(t)$ are linear functions of themselves, which is the major consequence of the Koopman representation. We now adopt a terse vector notation $\bfvg_\MM\l(t\r)=\l[g_1(t),\ldots g_M(t) \r]^T$ and $\bfvg_\MMbar\l(t\r)=\l[\bar{g}_1(t), \bar{g}_2(t) \ldots \r]^T$ and concisely express the full dynamics as:
\begin{equation} 
\frac{\dd}{\dd t} \begin{bmatrix}
\bfvg_\MM(t) \\
\bfvg_\MMbar(t) 
\end{bmatrix}
=\bfvL\cdot \begin{bmatrix}
\bfvg_\MM(t) \\
\bfvg_\MMbar(t) 
\end{bmatrix}
:=
 \begin{bmatrix}
\bfvL_{\MM\MM} &  \bfvL_{\MM\MMbar} \\
\bfvL_{\MMbar\MM} & \bfvL_{\MMbar\MMbar}\
\end{bmatrix}
\cdot
 \begin{bmatrix}
\bfvg_\MM(t) \\
\bfvg_\MMbar(t) 
\end{bmatrix}. \label{eq:fullKoopman}
\end{equation}
Here, the matrices $\bfvL_{i,j}$, $i,j \in \l\{\MM,\MMbar\r\}$, quantifies the effect from set $j$ to set $i$ in the linear evolution. While these matrices can be found explicitly from Eqs.~\eqref{eq:giEvol} and \eqref{eq:giBarEvol} should one know the Koopman eigenfunctions, we derive them just for the purpose of establishing the linear evolutionary equation \eqref{eq:fullKoopman}. We emphasize that although $g_i(0)=g_i \in \Hilg$ and $\bar{g}_i(0)=\bar{g}_i \in \Hilgbar$,  for $t>0$, $g_i(t)$ and $\bar{g}_i(t)$ are not necessarily in $\Hilg$ and $\Hilgbar$ respectively, if the interactions between the spaces, $\bfvL_{\MM\MMbar}$ and $\bfvL_{\MMbar\MM}$, are not zero. In other words, in general, both $g_i(t)$ and $\bar{g}_i(t)$ have nonzero components of the basis functions in $\Hilg$ and $\Hilgbar$. Nevertheless, the above linear equation \eqref{eq:fullKoopman} always holds.  

To obtain a closed-form evolution for our observables of interest in $\MM$, we first solve for the observables in set $\MMbar$  implicitly. Treating $\bfvL_{\MMbar\MM} \bfvg_\MM$ as an inhomogeneous driving term of the linear system, we solve the linear evolutionary equation for $\bfvg_\MMbar$:
\begin{align}
\bfvg_\MMbar(t) ={}&  \int_0^t e^{\l(t-s\r) \bfvL_{\MMbar\MMbar}} \cdot \bfvL_{\MMbar\MM}\cdot\bfvg_\MM(s) \dd s+e^{t \bfvL_{\MMbar\MMbar}} \cdot \bfvg_\MMbar(0). \label{eq:underResolvedDynamics}
\end{align}
The implicit solution of $\bfvg_\MMbar$ is in turn used to express closed evolutionary equations for the observables in the set $\MM$. 
\begin{align}
\frac{\dd}{\dd t} \bfvg_\MM(t) ={}& \bfvL_{\MM\MM} \bfvg_\MM\l(t\r) +\bfvL_{\MM\MMbar}\int_0^t e^{\l(t-s\r) \bfvL_{\MMbar\MMbar}} \cdot \bfvL_{\MMbar\MM}\cdot\bfvg_\MM(s) \dd s \label{eq:pGLE}  \\
{}& +\bfvL_{\MM\MMbar} e^{t \bfvL_{\MMbar\MMbar}} \cdot \bfvg_\MMbar(0).  \nonumber 
\end{align}
Equation \eqref{eq:pGLE} is almost closed in our chosen set of variables except for the last term. The first term is the instantaneous configuration of the set of observables applied to the physical-space configuration at time $t$ and the second is a delayed impact of the set of observables applied to the physical-space variables at an earlier time $s<t$. Both these terms depend only on the resolved observables $\bfvg_\MM$ at time $t$. However, the third term is induced by the initial setting of the under-resolved observables, $\bfvg_\MMbar(0)$, which cannot be generally resolved. Equation \eqref{eq:pGLE} is exact if one knows both $\bfvg_\MMbar(0)$ and $\bfvg_\MMbar(0)$, in which case, the system is fully resolved. Unfortunately, we do not have direct access to $\bfvg_\MMbar(0)$ as they are under-resolved observables, and one has to postulate their configurations in practice. 

This simple analysis illustrates the essential intuition of the Mori--Zwanzig formalism. Because we only resolve a set of observables $\bfvg_\MM(t)$ of the full dynamics, the impact from other observables $\bfvg_\MMbar(t)$ in Eq.~\eqref{eq:fullKoopman} cannot be directly accessed. Instead, we indirectly estimate the effect $\bfvL_{\MM\MMbar}\cdot \bfvg\l(t\r)$ from Eq.~\eqref{eq:underResolvedDynamics}, which contains two parts: the impact of resolved observables $\bfvg_\MM(s)$ at an earlier time $s$, and the initial conditions of the orthogonal observables $\bfvg_\MMbar$. The former characterizes the ``echo'' of the set of our interested observables to itself: at an earlier time $s$, these observables made an impact to the under-resolved observables $\bfvg_\MMbar$ (via $\bfvL_{\MMbar\MM}$), and such an impact propagates among the under-resolved observables $\bfvg_\MMbar$ for $t-s$ time via $e^{(t-s) \bfvL_{\MMbar \MMbar}}$ before coming back to affect the resolved observables at time $t$ via $L_{\MM\MMbar}$. The second part is a generic impact from the initial configuration of the orthogonal set of observables, $\bfvg_{\MMbar}(0)$, which has propagated in the under-resolved observables until the current time $t$ and affects the resolved observables. In the end, Eq.~\eqref{eq:pGLE} tells us that the accurate evolutionary equations of $\bfvg_\MM(t)$ always depend on (1) their instantaneous configuration, (2) their past history, and (3) an external ``driving force'' which depends on the initial configurations in the orthogonal space. 

We make a remark that the second and the third term are zero if the dynamics is closed in $\MM$, that is, $\bfvL_{\MM \MMbar}=0$ which corresponds to the scenario when we have a complete set of observables to describe the full dynamics. For example, this would be the $g_e:=\exp(-1/x)$ for the dynamics $\dot{x}=-x^2$. The memory and the external driven force exist only because we have an incomplete observable set in $\Hil$. 

We now drop the subscript $\MM$ in $\bfvg_\MM$, as we only care about the dynamics of the resolved observables. By defining an $M\times M$ matrix $\bfvM:=\bfvL_{\MM\MM}$, an $M\times M$ matrix parametrized by $\bfvK\l(s\r) :=- \bfvL_{\MM\MMbar} e^{s \bfvL_{\MMbar\MMbar}} \cdot \bfvL_{\MMbar\MM}$ parametrized by $s\in \mathbb{R}^+$, and an $M\times 1$ matrix $\bfvF(t):=\bfvL_{\MM\MMbar} e^{t \bfvL_{\MMbar\MMbar}} \cdot \bfvg_\MMbar(0)$, we arrive at the GLE
\begin{equation}
\frac{\dd}{\dd t} \bfvg\l(t\r) = \bfvM \cdot \bfvg\l(t\r) - \int_{0}^t \bfvK\l(t-s\r)\cdot \bfvg\l(s\r)\dd s + \bfvF\l(t\r).\label{eq:GLE}
\end{equation}
In the rest of the paper, we refer to $\bfvM$ as the \emph{Markov transition matrix}, $\bfvK(s)$ as the \emph{memory kernel}, and $\bfvF(t)$ as the \emph{orthogonal dynamics}. Although $\bfvF(t)$ is fully deterministic in Mori--Zwanzig formalism for deterministic systems, it is often referred to as the \emph{noise} because its resemblance of a Langevin noise in a Langevin equation. Note that the operators $\bfvM$ quantify the interactions within the group of the relevant observables $\l\{g_i\r\}_{i=1}^M$ (that is, $\bfvL_{\MM \MM}$). In contrast, the memory kernel combines the effects of the rest of the interactions ($\bfvL_{\MM \MMbar}$, $\bfvL_{\MMbar \MM}$, and $\bfvL_{\MMbar \MMbar}$)

Finally, we remark that despite their different terminologies, Eq.~\eqref{eq:GLE} and the operator algebraic formulation Eq.~\eqref{eq:opGLE} are equivalent. When $g_i(\bfv{x})=x_i$, given an initial condition $\bfvPhi_0$, $g_i\circ \bfvPhi_0$ in Eq.~\eqref{eq:GLE} \emph{is} $\phi_i \l(t, \bfvPhi_0\r)$ in Eq.~\eqref{eq:opGLE}. 

\subsection{Geometric interpretation of the GLE} \label{sec:geometry}

\begin{figure}[!t]
\centering
\includegraphics[width=0.8\textwidth]{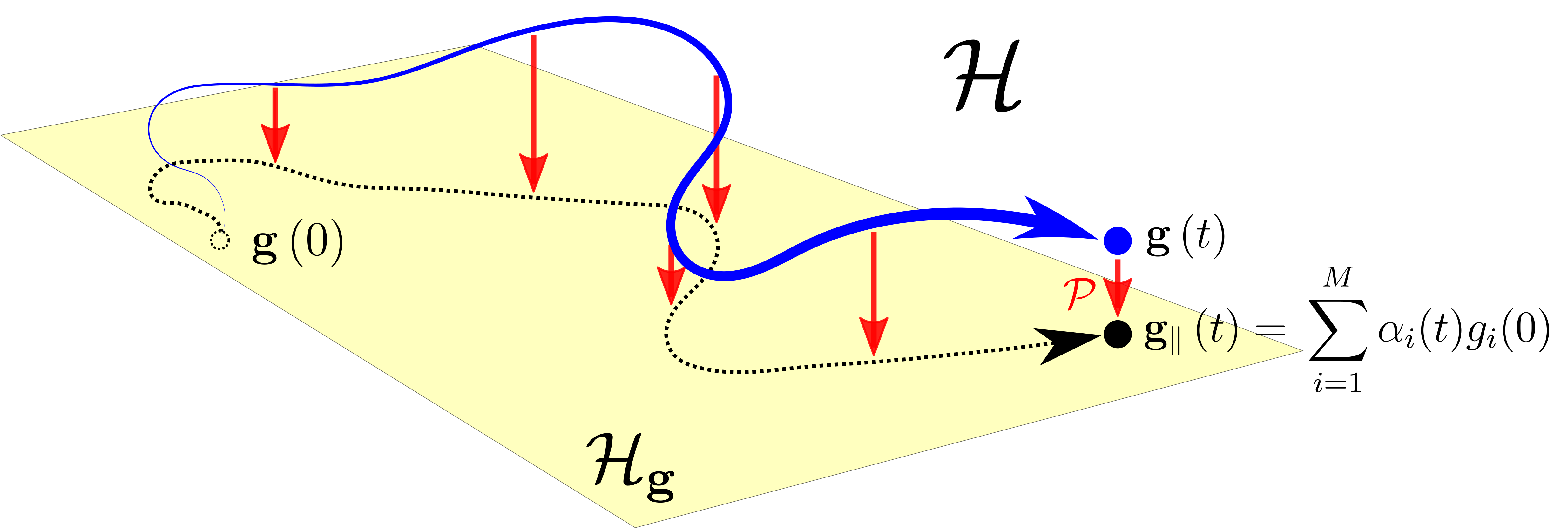}
\caption{Schematic diagram of the Mori--Zwanzig formalism. The Hilbert space $\Hil$ contains all possible observables which are functions of the initial conditions of the physical-space variables $\bfvPhi_0$. The subspace $\Hilg$ is linearly spanned by the set of selected observables $\text{Span}(\l\{g_i \r\}_{i=1}^M)$. We define the functions $g_i(t)$ parametrized by time $t$ to be $g_i(t) \circ \bfvPhi_0 :=g_i \circ \bfvPhi_t$, noting that $\bfvPhi(t)$ are functions of the initial condition $\bfvPhi_0$. At time $t=0$, the vector $\bfvg\l(0\r) = \l[g_1(0) , \ldots, g_M(0) \r]^T $ is exactly the vector of the basis $\l\{g_i \r\}_{i=1}^M$ and thus, $\bfvg\l(0\r)$ is in $\Hilg$. As the nonlinear dynamics evolves, $\bfvg\l(t\r)$ does not necessarily stays invariantly in $\Hilg$; in other words, $\bfvg\l(t\r)$ is not necessarily a linear combination of $\l\{g_i\r\}_{i=1}^M$. Mori's projection operator $\mathcal{P}$ projects $\bfvg(t)$ into $\Hilg$. The image is termed as $\bfvg_\parallel(t)$, which is a linear combination of $\l\{g_i \r\}_{i=1}^M$ and $\bfvg_\parallel(t)$ satisfies Eq.~\eqref{eq:PE}.}\label{fig:1}
\end{figure}

The GLE describe the exact evolution of $\bfvg(t)$ in $\Hil$. Figure \ref{fig:1} illustrates a schematic diagram of the dynamics of $\bfvg(t)$ in the space $\Hil$.
Because the GLE \eqref{eq:GLE} is linear, the observables $\bfvg(t)$ can be decomposed into two components: $\bfvg(t)=\bfvg_{\parallel}\l(t\r) + \bfvg_\perp \l(t\r)$. We define the parallel component $\bfvg_{\parallel}\l(t\r)$ as the general solution of the linear system and satisfies 
\begin{subequations}\label{eq:PE}
\begin{align}
\dot{\bfvg}_\parallel\l(t\r) ={}& \bfvM \cdot \bfvg_\parallel\l(t\r) - \int_{0}^t \bfvK\l(t-s\r)\cdot \bfvg_\parallel\l(s\r)\dd s, \\
\bfvg_\parallel\l(0\r)={}&\bfvg\l(0\r).
\end{align}
\end{subequations}
Because \eqref{eq:PE} is linear, it is clear that $\bfvg_{\parallel}(t)$ is just linear combination of the initial observables, i.e., we can always express $\bfvg_{\parallel}(t)=\sum_{i=1}^M \alpha_i(t) \bfvg_i(0)$ with some time-dependent coefficients $\alpha_i(t)$. Then, for any $t\ge 0$, $\l(\bfvg_\parallel\l(t\r)\r)_i  \in \Hilg$, $i=1\ldots M$. The orthogonal component $\bfvg_{\perp}\l(t\r)$ is the particular solution of the linear system with the driven force $\bfvF(t)$ and satisfies
\begin{subequations}\label{eq:OE}
\begin{align}
\dot{\bfvg}_\perp\l(t\r) ={}& \bfvM \cdot \bfvg_\perp\l(t\r) - \int_{0}^t \bfvK\l(t-s\r)\cdot \bfvg_\perp\l(s\r)\dd s + \bfvF(t), \\
\bfvg_\perp\l(0\r)={}&0.
\end{align}
\end{subequations}
Recall that the orthogonal dynamics $\l(\bfvF(t)\r)_i \in \Hilgbar$, $i=1\ldots M$. Then, again due to the linearity of \eqref{eq:OE}, the orthogonal component $\l(\bfvg_{\perp}(t)\r)_i \in \Hilgbar$, for any time $t\ge 0$. 

The geometric representation in Fig.~\ref{fig:1} illustrates the closure problem in the Hilbert space: to evolve the exact $\bfvg(t)$, one needs to provide the orthogonal dynamics $\bfvF(t)$. Solving $\bfvF(t)$ is as difficult as solving the full dynamics in the physical-space picture---we would need all the information of the orthogonal components, and thus the Mori--Zwanzig formalism has little to no advantage over the physical-space picture if the task is to solve for the exact $\bfvg(t)$. In fact, there is no free lunch by changing the Perron--Frobenius picture to the dual Koopman picture. To move forward, traditional analysis aim to select a set of slow-evolving observables---often referred to as the coarse-grained variables---and a noise model to replace the orthogonal dynamics $\bfvF(t)$. The rational of this approach is that, if the selected observables are complete to describe the slow dynamics, the orthogonal dynamics shall live at a much faster timescale and a proper noise model should suffice to \emph{model} the dynamics of $\bfvF(t)$. When a set of observables is given and when their dynamics when the timescales of the resolved and under-resolved observables are well-separated, Gottwald et al.~\cite{gottwald_crommelin_franzke_2017} provides a systematic slow-fast asymptotic analysis to homogenize effect of the orthogonal noise. The challenge is that it is not \emph{a priori} known what these observables are and what the corresponding noise model is, and identifying the observables and noise model is often from educated guesses supported by domain-specific knowledge.

\subsection{Mori's linear projection operator} \label{sec:projectionOperator}

Both the Koopman \cite{Koopman255,Koopman315,Mezic2005a} and Mori--Zwanzig formulations operate in a Hilbert space, whose inner product is commonly defined as the inner product of two functions as the expected value of the product of the two with respect to a chosen measure $\dd \mu$, Eq.~\eqref{eq:MoriInner}. Despite of the freedom to choose this measure, $\dd \mu$ is conventionally set as a natural measure that is specific to the dynamics. For example, it is natural to adopt the canonical equilibrium distribution (Gibbs' measure) for equilibrium Hamiltonian systems \cite{mori1965transport}. For non-equilibrium systems, we can adopt the stationary measure as $\dd \mu$ \cite{evans2007statistical}, as shown below. For stochastic and transient systems, it is natural to choose the induced time-dependent distribution \cite{wu2020VariationalApproachLearning}.

With a defined inner product, Mori used an operator $\mathcal{P}$ that projects any function (of the initial condition $\bfvPhi_0$) $f \in \Hil$, onto the subspace $\Hilg:=\text{span}\l(\MM\r)=\text{span}\l(\l\{g_i\r\}_{i=1}^M\r)$. As such, the projected image can be expressed by $\mathcal{P} f := \sum_{i=1}^M \alpha_i g_i$ with coefficients $\alpha_i$. Now, we decompose the function $f$ by $ f= \mathcal{P} f + f_\perp$ where $f_\perp \in \Hilgbar$. Because $\l\langle f_\perp, g_j \r\rangle=0$, $j=1\ldots M$, the inner product $\l\langle f, g_j \r\rangle$ is:
\begin{equation}
\l\langle f, g_j \r\rangle = \l\langle \mathcal{P} f, g_j \r\rangle  = \sum_{i=1}^M \alpha_i \l\langle g_i, g_j \r\rangle.  \label{eq:innerProduct}
\end{equation}
In vector notation, $\bfvg:=\l[g_1 \ldots g_M\r]^T$ and $\bfv{\alpha}:=\l[\alpha_1 \ldots \alpha_M\r]$, the above equation can be expressed concisely by $\l\langle f, \bfvg^T \r\rangle= \bfv{\alpha} \cdot \l\langle \bfvg, \bfvg^T \r\rangle$. Because the basis functions $g_i$'s are linearly independent, $\l\langle \bfvg , \bfvg^T \r\rangle$ is a full-rank and invertible matrix. Then, $\bfv{\alpha}=\l\langle f, \bfvg^T \r\rangle \cdot  \l\langle \bfvg , \bfvg^T \r\rangle^{-1}$, and we obtain the final expression for the projection operator
\begin{equation}
\mathcal{P} f = \l\langle f, \bfvg^T \r\rangle \cdot  \l\langle \bfvg , \bfvg^T \r\rangle^{-1} \cdot \bfvg.
\end{equation}
Note that if and only if the pre-selected set of functions $\l\{g_i\r\}_{i=1}^M$ are orthonormal with respect to the defined inner product, $ \l\langle \bfvg , \bfvg^T \r\rangle^{-1} $ is an identity matrix and the expression can be simplified by $\mathcal{P} f=\l\langle f, \bfvg^T \r\rangle   \cdot \bfvg$. Finally, we emphasize that the projection operator crucially depends on the choice of the defined inner product. In the rest of this article, we consider an inner product defined by averaging the observables over a long trajectory of the dynamical system:
\begin{equation}
\l\langle f , g\r\rangle := \lim_{T\rightarrow \infty}\frac{1}{T} \int_0^T \l(f\circ \bfvPhi(s) \r) \l( g\circ \bfvPhi(s)\r) \dd s, \label{eq:ctsInnerProduct}
\end{equation}
where $\bfvPhi\l(t\r)$ is the solution of Eq.~\eqref{eq:phiEvo} from any initial condition, assuming the choice of the initial condition does not change the long-time statistics. 

\subsection{The evolution of the projected image and time correlation matrix}\label{sec:CE+OP}
The GLE \eqref{eq:GLE} is not a closed dynamical system for the exact $\bfvg(t)$ because it contains the generalized Langevin noise, $\bfvF$, which is not known and hard to obtain. However, because $\bfvg(t)=\bfvg_\parallel + \bfvg_\perp$, the projected image $\mathcal{P}\bfvg(t)=\mathcal{P}\bfvg_\parallel(t)+\mathcal{P}\bfvg_\perp(t)=\bfvg_\parallel$ follows the closed evolutionary Eq.~\eqref{eq:PE} and with an initial condition $\mathcal{P}\bfvg(0)=\bfvg(0)$. The $L^2$-norm of the residual error of the projected image $\Vert\bfvg - \mathcal{P}\bfvg \Vert_2^2 := \l\langle\bfvg - \mathcal{P}\bfvg , \bfvg - \mathcal{P}\bfvg \r\rangle $ is minimal among all the schemes that decompose $\bfvg(t)$ into a \emph{parallel} component and a \emph{non-parallel} component. Thus, the projected image $\mathcal{P}\bfvg(t)$ is the best approximation to  the evolution of g(t) in the parallel space, and in this sense one can conceive it as the optimal predictor of the exact evolution $\bfvg(t)$ into the future ($t>0$). Somewhat interestingly, this shows that the optimal prediction into the future defined as above only depends on the choice of the inner product (and consequently the projection operator) and does not depend on the orthogonal dynamics $\bfvF(t)$. 

A related way to close the dynamics is to apply $\l\langle \cdot, \bfvg^T \r\rangle$ to the GLE \eqref{eq:GLE}, resulting in the evolutionary equation for the two-time correlation function $\bfvC(t):=\l\langle \bfvg(t), \bfvg^T \r\rangle$:
\begin{equation}
\frac{\dd}{\dd t} \bfvC(t) = \bfvM \cdot \bfvC(t) - \int_0^t \bfvK\l(t-s\r) \bfvC(s) \dd s, \quad \bfvC(0) = \l\langle  \bfvg, \bfvg^T \r\rangle. \label{eq:CE}
\end{equation}
Here, $\bfvC(t)$ is the expected two-time correlation of the observables with respect to an initial condition $\bfvPhi_0$ distributed according to a long-time statistics $\dd \mu$ (cf.~Eq.~\eqref{eq:ctsInnerProduct}). 
Multiplying $\bfvC^{-1}(0) \cdot \bfvg(0)$ to Eq.~\eqref{eq:CE} from the right, we obtain
\begin{equation}
\frac{\dd}{\dd t} \bfvC(t)\cdot\bfvC^{-1}(0) \cdot \bfvg(0) = \bfvM \cdot \bfvC(t)\cdot \bfvC^{-1}(0) \cdot \bfvg(0) - \int_0^t \bfvK\l(t-s\r) \bfvC(s)\cdot \bfvC^{-1}(0) \cdot \bfvg(0) \dd s. \label{eq:15210121}
\end{equation}
Comparing Eq.~\eqref{eq:15210121} to the evolutionary Eq.~\eqref{eq:PE}, we immediately identify the solution of the projected image $\bfvg_\parallel(t)$:
\begin{equation}
\bfvg_{\parallel}(t) = \bfvC(t) \cdot \bfvC^{-1}(0) \cdot \bfvg(0).\label{eq:OP}
\end{equation}
Thus, the temporal correlation matrix $\bfvC(t)$ encodes the information for optimally predicting the dynamics into the future. In addition, it is straightforward to solve the orthogonal component $\bfvg_\perp(t)$ as a superposition of the orthogonal driven force $\bfvF(t)$:
\begin{equation}
\bfvg_\perp(t)=\int_0^t  \bfvC(t-s)\cdot \bfvC^{-1}(0) \cdot \bfvF(s) \dd s.  \label{eq:gperpsolution}
\end{equation}
We illustrate the analysis in Appendix \ref{app:orthogonal}. 

As will be seen in Sec.~\ref{sec:algos}, Eq.~\eqref{eq:CE} plays a pivotal role for data-driven learning of the Markov transition $\bfvM$ and memory kernel $\bfvK(s)$, $s\ge 0$.
We will also see that the solution of the optimal prediction Eq.~\eqref{eq:OP} establishes the equivalence between the Mori--Zwanzig formalism and the approximate Koopman learning algorithm in Sec.~\ref{sec:koopman}.

\subsection{Generalized Fluctuation-Dissipation Relationship} \label{sec:GFD}
With a suitable choice of the inner product, there exists a subtle relationship---often referred to as the Generalized Fluctuation-Dissipation (GFD) relationship---between the memory kernel $\bfvK$ and the orthogonal dynamics $\bfvF$: 
\begin{equation}
\bfvK\l( s \r) = \l\langle \bfvF(s), \bfvF^T(0) \r\rangle \bfvC^{-1}(0). \label{eq:GFD}
\end{equation}
Using the notations defined in Sec.~\ref{sec:intuitiveDerivation}, we illustrate how this subtle relationship emerges.
Explicitly, from Eq.~\eqref{eq:pGLE}, we identify the orthogonal dynamics
\begin{align}
\bfvF(t) = \bfvL_{\MM \MMbar} e^{t \bfvL_{\MMbar \MMbar}} \bfvg_\MMbar,
\end{align}
because $\bfvg_\MMbar(0)\equiv \bfvg_\MMbar$. 
Next, with the choice of the temporal averaging inner product Eq.~\eqref{eq:ctsInnerProduct}, we can obtain:
\begin{align}
\l\langle \bfvF(t), \bfvF^T(0) \r\rangle = \lim_{T\rightarrow \infty} \frac{1}{T} \int_0^T \bfvL_{\MM \MMbar} e^{t \bfvL_{\MMbar \MMbar}} \bfvg_\MMbar\circ\bfvPhi(s) \cdot \l[ \bfvL_{\MM \MMbar} \bfvg_\MMbar\circ\bfvPhi(s) \r]^T \dd s. \label{eq:GFD-1}
\end{align}
Induced by the dynamics Eq.~\eqref{eq:phiEvo}, $\bfvg_\MMbar (s)\equiv \bfvg_\MMbar \circ \bfvPhi(s)$ and $\bfvg_\MM (s)\equiv  \bfvg_\MM \circ \bfvPhi(s)$ satisfy Eq.~\eqref{eq:fullKoopman}, so we use the identity
\begin{equation}
\bfvL_{\MM \MMbar} \bfvg_\MMbar (s) = \frac{\dd}{\dd s} \bfvg_\MM(s) - \bfvL_{\MM \MM} \bfvg_\MM (s)
\end{equation}
to replace $\bfvL_{\MM \MMbar} \bfvg_\MMbar (s)$ in \eqref{eq:GFD-1}:
\begin{align}
\l\langle \bfvF(t), \bfvF^T(0) \r\rangle={}& \lim_{T\rightarrow \infty} \frac{1}{T} \int_0^T \bfvL_{\MM \MMbar} e^{t \bfvL_{\MMbar \MMbar}} \bfvg_\MMbar (s) \cdot \l[\frac{\dd}{\dd s} \bfvg_\MM(s) - \bfvL_{\MM \MM} \bfvg_\MM (s) \r]^T \dd s  \nonumber\\
={}& \lim_{T\rightarrow \infty} \frac{1}{T} \int_0^T \bfvL_{\MM \MMbar} e^{t \bfvL_{\MMbar \MMbar}} \bfvg_\MMbar (s) \cdot  \frac{\dd}{\dd s} \bfvg_\MM^T(s) \dd s   \nonumber \\{}& - \bfvL_{\MM \MMbar} e^{t \bfvL_{\MMbar \MMbar}} \l\langle \bfvg_\MMbar  , \bfvg_\MM^T  \r\rangle  \bfvL_{\MM \MM}^T \nonumber \\
={}& \lim_{T\rightarrow \infty} \frac{1}{T} \int_0^T \bfvL_{\MM \MMbar} e^{t \bfvL_{\MMbar \MMbar}} \bfvg_\MMbar (s) \cdot  \frac{\dd}{\dd s} \bfvg_\MM^T(s) \dd s. 
\end{align}
In the last line, we have used the fact that $\bfvg_\MM$ and $\bfvg_\MMbar$ are orthogonal with respect to the inner product by construction. Next, we perform an integration by part, assume that the boundary terms are bounded and thus converge to $0$ after divided by $T$ in the limit $T\rightarrow \infty$, and use the full dynamics Eq.~\eqref{eq:fullKoopman} again to obtain 
\begin{align}
\l\langle \bfvF(t), \bfvF^T(0) \r\rangle ={}& - \lim_{T\rightarrow \infty} \frac{1}{T} \int_0^T \bfvL_{\MM \MMbar} e^{t \bfvL_{\MMbar \MMbar}} \frac{\dd}{\dd s}  \bfvg_\MMbar (s) \cdot   \bfvg_\MM^T(s) \dd s \nonumber \\
={}&  - \lim_{T\rightarrow \infty} \frac{1}{T} \int_0^T \bfvL_{\MM \MMbar} e^{t \bfvL_{\MMbar \MMbar}} \l[\bfvL_{\MMbar \MM} \bfvg_{\MM}(s)+\bfvL_{\MMbar \MMbar} \bfvg_{\MMbar}(s)\r] \cdot   \bfvg_\MM^T(s) \dd s \nonumber \\
={}&  - \bfvL_{\MM \MMbar} e^{t \bfvL_{\MMbar \MMbar}} \l[\bfvL_{\MMbar \MM}  \l\langle \bfvg_{\MM}, \bfvg^T_{\MM} \r\rangle + \bfvL_{\MMbar \MMbar}  \l\langle \bfvg_{\MMbar}, \bfvg^T_{\MM} \r\rangle\r] \nonumber\\
={}& - \bfvL_{\MM \MMbar} e^{t \bfvL_{\MMbar \MMbar}} \bfvL_{\MMbar \MM}  \l\langle \bfvg_{\MM}, \bfvg^T_{\MM} \r\rangle.
\end{align}
Again, we used the orthogonality $\l\langle \bfvg_\MM,  \bfvg_\MMbar \r\rangle=0$. Finally, the memory kernel $\bfvK(t):= - \bfvL_{\MM \MMbar} e^{t \bfvL_{\MMbar \MMbar}} \bfvL_{\MMbar \MM}$ can be expressed as the two-time correlation statistics of the orthogonal dynamics $\bfvF(t)$ and the autocorrelation of the observables $  \l\langle \bfvg_{\MM}, \bfvg^T_{\MM} \r\rangle$:
\begin{equation}
\bfvK(t)= \l\langle \bfvF(t), \bfvF^T(0) \r\rangle \cdot   \l\langle \bfvg_{\MM}, \bfvg^T_{\MM} \r\rangle^{-1}.
\end{equation}

The above relationship between the two-time statistics of the orthogonal dynamics $\bfvF(t)$ and the memory kernel $\bfvK(t)$ is referred to as the Generalized Fluctuation-Dissipation relationship. In the operator algebraic derivation (cf.~\ref{sec:operatorAlgebra}), GFD holds when the Liouville operator $\mathcal{L}=\mathcal{L}\l(\Phi\r)$ is anti self-adjoint with respect to the chosen inner product, i.e., for any test functions $f$ and $h$ of the physical-space variable $\bfvPhi$,
\begin{equation}
\l\langle f, \mathcal{L} h \r\rangle = -\l\langle \mathcal{L}  f, h \r\rangle. \label{eq:antiSelfAdjointness}
\end{equation}
For Hamiltonian systems, the anti self-adjointness is guaranteed directly from the volume-preserving property of the dynamics \cite{Koopman255}. For non-equilibrium systems, using long time-averaging as the inner product also has the anti self-adjoint property: 
\begin{align}
\l\langle f, \mathcal{L} h \r\rangle ={}& \l\langle f, \frac{\dd}{\dd t} h \r\rangle = \lim_{T\rightarrow \infty}\frac{1}{T} \int_0^T f\circ \bfvPhi(t) \frac{\dd}{\dd t} \l[ g\circ \bfvPhi(t)\r]\dd t \nonumber \\
={}& - \lim_{T\rightarrow \infty}\frac{1}{T} \int_0^T \frac{\dd}{\dd t} \l[f\circ \bfvPhi(t)\r]  g\circ \bfvPhi(t) \dd t = -\l\langle \frac{\dd}{\dd t} f, h \r\rangle = -\l\langle \mathcal{L}  f, h \r\rangle.
\end{align}
All we need are the minor conditions that the integration by part is valid, and negligible boundary terms which can be guaranteed for bounded systems.

Furthermore, if the dynamical system is ergodic, the temporal average \eqref{eq:ctsInnerProduct} converges to \eqref{eq:MoriInner}, in which $\dd \mu =\rho_\text{stat} \l(\bfvPhi_0\r) \dd^N \bfv{x}$ where $\rho_\text{stat}$ is the stationary density function and satisfies $e^{t\mathcal{L}^*}  \rho_\text{stat} = \rho_\text{stat}$; here, the adjoint $\mathcal{L}^\ast$ defines the Perron-Frobenius operator. In the literature, it is generally presented that the anti self-adjointness is valid for Hamiltonian systems in a heat bath, e.g., where there is an induced Gibbs measure \cite{AlexandreJ.Chorin2002}. Our analysis shows that GFD is generally valid, even for non-equilibrium (not necessarily Hamiltonian) systems, so long as we choose time-averaging as the inner product and the observables along a long trajectory is bounded. 

Finally, GFD can be considered as a self-consistent condition of the memory kernel $\bfvK$ and the orthogonal dynamics $\bfvF$. A real stochastic Langevin system with a white noise (no time correlation) would result in a $\bfvK(t)=\bfvK_0 \, \delta\l(t\r)$ where $\bfvK_0$ is a constant matrix and $\delta$ is the Dirac $\delta$-distribution. The Mori--Zwanzig formalism shows that when a dynamical system is not fully resolved, in general, there exists a non-zero memory kernel $\bfvK$ and thus the self-consistent orthogonal dynamics $\bfvF$ must be a color noise. Below in Sec.~\ref{sec:algos} we provide data-driven algorithms to extract the memory kernel and the noise directly from the measured observables along a long trajectory. The GFD can serve as a very stringent self-consistency check for the algorithms.  

\subsection{A discrete-time Mori--Zwanzig formalism} \label{sec:dMZ}
Even though we are interested in a continuous-time model, it is common that the observations---either empirical measurements of the physical system or the output of computer simulations---are discrete in time. It is also desirable to store the trajectories of a large set of observables sparsely sampled in time to mitigate to the storage limitation. In this case, the continuous-time Mori--Zwanzig formalism is not adequate. In this section, we provide the result of a discretized Mori--Zwanzig formulation that is more suitable for discrete-time data. For brevity, we only present the result in this section and leave the tedious yet straightforward derivation in Appendix \ref{app:discrete-time}.

We consider to observe the continuous-time system at discretization of times, $t=k\Delta$, $k\in \mathbb{Z}_{\ge_0}$ and $\Delta$ is not necessarily small. After integrating the GLE Eq.~\eqref{eq:GLE} and the evolutionary equations for the correlation matrix $\bfvC(t) $ (Eq.~\eqref{eq:CE}) and the optimal prediction $\mathcal{P}\bfvg\l(t\r)$  (Eq.~\eqref{eq:OP}) at the discrete times, the snapshots of the observables of interests satisfy very similar mathematical structures of the continuous-time formulations. Specifically, in Appendix \ref{app:discrete-time-1}, we establish the discrete-time GLE (cf. Eq.~\eqref{eq:GLE})
\begin{equation}
\bfvg\l( \l(k+1\r)\Delta \r)  =   \sum_{\ell=0}^{k} \bfvO_\Delta^{\l(\ell\r)} \cdot  \bfvg(\l(k-\ell\r) \Delta) + \bfvW_{k}, \label{eq:dGLE}
\end{equation}
where $\bfvO_\Delta^{\l(\ell\r)}$'s are $M\times M$, $\Delta$-dependent matrices which can be defined in terms of the continuous-time Markov transition matrix $\bfvM$ and memory kernel $\bfvK$, and $\bfvW_{k}$ is the discrete-time orthogonal dynamics which can be decomposed into linear functions of the continuous-time orthogonal dynamics $\bfvF$. Similarly, the snapshots of the correlation matrix $\bfvC$ satisfy (cf. Eq.~\eqref{eq:CE})
\begin{equation}
\bfvC\l( \l(k+1\r)\Delta \r)  =   \sum_{\ell=0}^{k} \bfvO_\Delta^{\l(\ell\r)} \cdot \bfvC(\l(k-\ell\r) \Delta),\label{eq:dCE}
\end{equation}
and the snapshots of the projected image $\mathcal{P}\bfvg(t)$ satisfy (cf. Eq.~\eqref{eq:OP})
\begin{equation}
\mathcal{P}\bfvg\l( \l(k+1\r)\Delta \r)  =   \sum_{\ell=0}^{k} \bfvO_\Delta^{\l(\ell\r)} \cdot \mathcal{P} \bfvg(\l(k-\ell\r) \Delta). \label{eq:dOP}
\end{equation}
It is tempting to associate the operator $\bfvO_\Delta^{\l(0\r)}$ to the continuous-time Markov matrix $\bfvO_\Delta^{\l(0\r)} \approx \bfv{I} +\bfvM \Delta$, where $\bfv{I}$ is the $M\times M$ identity matrix, and to associate the operator $\bfvO_\Delta^{\l(\ell\r)}$, $\ell>1$, to the continuous-time memory kernel by $\bfvO_\Delta^{\l(\ell\r)} \approx \Delta^2 \bfvK\l(\ell \Delta \r)$. We remark that these representations involving the infinitesimal $\Delta \ll 1$ are mathematically valid, but in general for finite $\Delta$, the expressions of the operators $\bfvO_\Delta^{\l(\ell\r)}$ in terms of the continuous-time objects ($\bfvM$ and $\bfvK(s)$) are not simple; see Lemma \ref{lemma:1} in Appendix \ref{app:discrete-time-1}. Nevertheless, the above Eqs.~\eqref{eq:dGLE}, \eqref{eq:dCE}, and \eqref{eq:dOP} are always valid regardless of the choice of $\Delta$. We also establish the GFD relationship in Appendix \ref{app:discrete-time-GFD}.

For a generic discrete-time dynamical system, it is also possible to directly derive its Mori--Zwanzig formula (Appendix \ref{app:discrete-time-2} and reference \cite{Lin2021}). We remark that while the mathematical expression of the results of this approach look identical to Eqs.~\eqref{eq:dGLE}-\eqref{eq:dOP}, there exists a subtlety between discretizing the correlation matrix $\bfvC(t)$ of a continuous-time system (Appendix \ref{app:discrete-time-1}) and the generic discrete-time correlation matrix (Appendix \ref{app:discrete-time-2}). We discuss such a subtlety in Appendix~\ref{app:discrete-time-diff}. Furthermore, in Appendix \ref{app:discrete-time-GFD-2}, we show that the GFD relationship is valid when the discrete-time operator $\mathcal{L}_d$, which forward propagates the observables (i.e., $\bfvg\l(\l(k+1\r)\Delta \r)= \mathcal{L}_d \bfvg\l(k\Delta \r)$), is anti-self-adjoint with respect to the invariant measure of the discrete-time dynamics (cf. Sec.~\ref{sec:GFD}). 

\section{Relation to the approximate Koopman learning methods} \label{sec:koopman}
The discretized formulations presented in Sec.~\ref{sec:dMZ} provide the mathematical structure for making comparison to the approximate Koopman learning framework \cite{schmid2010DMD,schmid2011applications,williams2015data}. In this section, we establish that the Mori--Zwanzig formalism with Mori's projection operator is not only compatible with, but also a generalization of the Koopman learning framework. 

We begin with a short summary of the approximate Kooman learning methods, specifically, the extended dynamic mode decomposition (EDMD, \cite{williams2015data}). EDMD takes a long trajectory of a set of $M$ observables, $\l\{g_i\r\}_{i=1}^M$, of a nonlinear dynamical system as an input. The snapshots of these descriptors on a uniformly separated temporal grid were recorded as $g_i\l(j\delta\r)$, $j=0\ldots N-1$. Generally, $\delta$ is conceived as a small time separation. The goal of the EDMD is to identify the \emph{approximate Kooman operator} $\bfv{{K}}_\delta^{\text{Koop}}$---note the unfortunate convention of $\bfvK$ for memory kernel in the context of Mori--Zwanzig---which linearly maps the previous snapshots to the consecutive next ones with a minimal squared residual error. Operationally, EDMD stacks up the snapshots into an array of ``dependent variables'' $\bfv{Y}$ and ``independent variables'' $\bfv{X}$:
\begin{align}
\bfv{Y} =
\begin{bmatrix}
\bfvg_1\l(1 \delta \r) & \ldots  & \bfvg_1\l(\l(N-1\r) \delta \r) \\
\bfvg_2\l(1 \delta \r) & \ldots  & \bfvg_2\l(\l(N-1\r) \delta \r) \\
\vdots  &  \ddots & \vdots \\
\bfvg_M\l(1 \delta \r)  & \ldots  & \bfvg_M\l(\l(N-1\r) \delta \r) \\
\end{bmatrix},\quad 
\bfv{X} =
\begin{bmatrix}
\bfvg_1\l(0 \delta \r) & \ldots  & \bfvg_1\l(\l(N-2\r) \delta \r) \\
\bfvg_2\l(0 \delta \r) & \ldots  & \bfvg_2\l(\l(N-2\r) \delta \r) \\
\vdots & \ddots & \vdots \\
\bfvg_M\l(0 \delta \r) & \ldots  & \bfvg_M\l(\l(N-2\r) \delta \r) \\
\end{bmatrix},
\end{align}
and the $M\times M$ matrix $\bfv{\hat{K}}_\delta^{\text{Koop}}$ is the linear operator which minimizes the mean squared error
\begin{equation}
\varepsilon^2 := \frac{1}{N-1} \sum_{j=1}^{N-1}\sum_{i=1}^{M}\l(\bfv{y} - \bfv{{K}}_\delta^{\text{Koop}} \cdot \bfv{x}\r)_{i,j}^2.
\end{equation}
When the observables form a complete set of basis functions, the dynamics is closed in the linear spanned space $\Hilg$, and $\varepsilon^2$ can be minimized to zero. Nevertheless, for general problems, it is challenging to identify a complete set of basis functions. Consequently, there exist non-zero residuals. The learning problem is fundamentally a linear regression problem, which is concave. Formally, the unique minimizer $ \bfv{\hat{K}}_\delta^{\text{Koop}}$ conditioned on a pair of $\bfv{Y}$ and $\bfv{X}$ is
\begin{equation}
\bfv{{K}}_\delta^{\text{Koop}} = \l(\bfv{Y}\cdot \bfv{X}^T\r) \cdot \l(\bfv{X}\cdot \bfv{X}^T\r)^{-1},  \label{eq:EDMD}
\end{equation}
when the set of the observables are linearly independent. When the set of the observables are not linearly independent, the problem is under-determined and there exist a family of minimizers \cite{williams2015data}; in such a case, the inverse matrix $ \l(\bfv{X}\cdot \bfv{X}^T\r)^{-1}$ can be operationally carried out by the Moore--Penrose pseudoinverse \cite{williams2015data} to identify one of the minimizers. In the analysis below, we assume that the set of the basis functions is carefully chosen so that they are linearly independent.

Equation \eqref{eq:EDMD} exhibits the exact mathematical expressions of the projected image $\bfvg_\parallel(t)$ in the continuous-time Mori--Zwanzig formalism (Eq.~\eqref{eq:OP}).  In the limit of infinitely long snapshots separated by $\delta$, the matrices $\bfv{Y}\cdot\bfv{X}^T$ and $\bfv{X}\cdot\bfv{X}^T$ are exactly $\bfvC(\delta)$ and $\bfvC(0)$ where the inner product is defined to be with respect to the distribution induced by the long trajectories:
\begin{subequations}
\begin{align}
\bfvC(0) ={}& \l\langle \bfvg, \bfvg^T \r\rangle = \lim_{T\rightarrow \infty} \frac{1}{T} \int_{0}^T \bfvg\circ\bfvPhi\l(s\r)\cdot\bfvg^T\circ\bfvPhi\l(s\r) \dd s \approx \bfv{X}\cdot\bfv{X}^T, \\
\bfvC(\delta) ={}& \l\langle e^{\delta \mathcal{L}} \bfvg , \bfvg^T \r\rangle = \lim_{T\rightarrow \infty} \frac{1}{T} \int_{0}^T \bfvg\circ\bfvPhi\l(s+\delta\r)\cdot\bfvg^T\circ\bfvPhi\l(s\r) \dd s \approx \bfv{Y}\cdot\bfv{X}^T.
\end{align}
\end{subequations}
Thus, both formulations predict the exact propagator forward $\delta$-time $\bfvC(\delta) \cdot \bfvC^{-1}(0)$. 

It is intriguing that the Mori--Zwanzig formulation relies on the projection operator $\mathcal{P}$ which requires the equipped inner product of the Hilbert space, but the approximate Koopman learning framework only relies on the mean $L^2$-norm of the error which relies on a less strict norm space. Nevertheless, operationally, we can use the geometric interpretation provided in Sec.~\ref{sec:geometry} to illustrate the intuition of the two formulations. The Koopman learning framework seeks a point in $\Hilg$ that minimizes the $L^2$-norm between the point and $\bfvg(\delta)$, which is generally outside $\Hilg$ (see Fig.~\ref{fig:1}). In contrast, the Mori--Zwanzig formalism simply projects $\bfvg(\delta)$ onto $\Hilg$. These two formulations are formally identical because the projected image $\mathcal{P}\bfvg(\delta)$ would be the unique point on $\Hilg$ such that the $L^2$-error of the projected image to $\bfvg(\delta)$, $\left \Vert \bfvg(\delta)- \mathcal{P}\bfvg(\delta)\right \Vert_2^2 =\l\langle \bfvg(\delta)- \mathcal{P}\bfvg(\delta), \bfvg(\delta)- \mathcal{P}\bfvg(\delta) \r\rangle$ is minimized. We remark the subtle difference between the two frameworks. In the Koopman framework, orthogonality between the residual and the $\Hilg$ is not established---there is no notion of an inner product. In the Mori--Zwanzig formalism with the equipped inner product in $\Hil$, the unique operator $\bfvC(k \delta)\cdot \bfvC^{-1}(0)$, $k\in \mathbb{N}$ propagates $\bfvg(0)$ to $\bfvC(k \delta)\cdot \bfvC^{-1}(0)\cdot \bfvg(0)$ which is \emph{always} the projected image of $\bfvg(k\delta)$, and the residual $\bfvg(k\delta)-\mathcal{P}\bfvg(k\delta)$ is \emph{always} orthogonal to $\Hilg$, with respect to the induced inner product. We remark that $\bfvC(\delta)\cdot \bfvC^{-1}(0)$ can also be derived from Rayleigh–-Ritz variational principle of the leading eigenvalues of either the Perron--Frobenius of Koopman operators using related algorithms such as the time-lagged independent component analysis and Algorithm for Multiple Unknown Signals Extraction, commonly used and historically founded in the Perron--Frobenius picture by the molecular dynamics community \cite{tong1990AMUSENewBlind,molgedeySeparationMixtureIndependent1994,perez-hernandez2013IdentificationSlowMolecular,noe2013VariationalApproachModeling,nuske2014VariationalApproachMolecular,wu2017VariationalKoopmanModels,klus2018DataDrivenModelReduction,wu2020VariationalApproachLearning}. 

We have established the equivalence of the Mori--Zwanzig formalism and the approximate Koopman learning framework. Mori--Zwanzig is more general than the existing Koopman learning methods. First, Eq.~\eqref{eq:dOP} prescribes the optimal prediction $\mathcal{P}\bfvg(\Delta) = \bfvO_{\Delta}^{(0)}\bfvg(0)$ if $\bfvg(0)$ was sampled from the measure $\dd \mu$ which was used to define the inner product, and the horizon $\Delta$ does not have to be small. Our analysis in Appendix \ref{app:discrete-time-1} shows that it is always possible to identify the unique (but $\Delta$-dependent) operator $\bfvO_{\Delta}^{(0)}$, regardless of how large $\Delta$ is. Interestingly, in our derivation provided in Appendix \ref{app:discrete-time-1}, we can see that $\bfvO_\Delta^{(0)}$ depends on not only the Markov transition $\bfvM$ but also the memory kernel $\bfvK(s)$, $s\in[0,\Delta)$. This formally states that when the system is not fully resolved, the forward operator cannot be simply approximated by $e^{t\bfvM}$. Instead, the linear operator $\bfvO_\Delta^{(0)}$, which can be estimated by our proposed algorithm in Sec.~\ref{sec:algos}, has an implicit memory-kernel ($\bfvK(s)$, $s\in\l[0,\Delta\r)$) dependence, and thus we cannot simply exponentiate it for predicting the system further than $\Delta$ into the future. Secondly, the discretized Generalized Langevin Equation \eqref{eq:dGLE} states that the prediction shall be made with the past history, when it is available and when the discretized memory kernel $\bfvO_{\Delta}^{(\ell)}$, $\ell\ge 1$ is not zero. By taking into the account of the past history, we will be simultaneously use the information in the parallel component $\bfvg_\parallel(j\Delta)$ and the perpendicular component $\bfvg_\perp(j\Delta)$ to forward propagate the system and thus reduce the prediction error. In contrast, in approximate Koopman learning, we only use the current time to predict the next time step and neglect the orthogonal contribution which would be estimated by the past trajectory in the context of Mori--Zwanzig formalism. The advantage of of the Mori--Zwanzig's history-dependent prediction will be numerically illustrated in Sec.~\ref{sec:errorAnalysis}.

\section{Learning the operators in the Mori--Zwanzig formalism from data} \label{sec:algos}
Conventionally, the aim of Mori--Zwanzig analysis is to select a set of coarse-grained variables to construct a parallel space $\Hilg$ in which the projected image $\mathcal{P}\bfvg(t)$ captures the slow modes of the dynamics, mostly described by the Markov transition matrix $\bfvM$. If the parallel space fully capture the slow modes, the orthogonal dynamics would predominantly be the fast modes of the dynamics. In such construction by time-scale separation, the orthogonal dynamics $\bfvF(t)$ are usually modeled by a generic stochastic process, and the self-consistent memory kernel can be constructed to establish the Generalized Langevin Equation \eqref{eq:GLE} describing the slow modes of the dynamics \cite{stinis2004stochastic,horenko07data,chen14computation,li2015incorporation,lei2016data,li2017computing,parish2017non,wang2019implicit,hudson20coarse}. 
 
The conventional approach is challenging because the selection of the observables requires sophisticated understanding of how to separate the modes with different timescales in a complex dynamical systems. Nevertheless, the GLE \eqref{eq:GLE} is always mathematically correct, and thus, if we have collected a large enough set of data of a dynamical system, it is possible to numerically estimate Markov transition matrix $\bfvM$, memory kernel $\bfvK$, and the orthogonal dynamics $\bfvF$ directly from the collected data. Here, we provide two algorithms, one for continuous-time models (Algorithm \ref{alg:1}) and the other for discrete-time models (Algorithm \ref{alg:2}) for estimating the key quantities in the Mori--Zwanzig formalism directly from long trajectories of the observables.  We remark that the scope of this study is restricted to noiseless observations of  deterministic systems. For systems with some form of randomness, such as noisy observation on a deterministic system or a generic stochastic system, the estimation of stochastic Koopman operators \cite{Mezic2005a,williams2015data,hudson20coarse,wu2017VariationalKoopmanModels} requires additional mathematical construction (a probability space) that is beyond the scope of our theoretical analysis in the previous sections. As such, we will defer these cases to future studies.

The procedure begins with calculating the two-time correlation functions $\bfvC(t)$ of the observables from the recorded long trajectory. Then, we exploit Eqs.~\eqref{eq:CE} for continuous-time systems and \eqref{eq:dCE} for discrete-time systems to estimate continuous-time $\bfvM$ and $\bfvK$ and various orders of discrete-time $\bfvO^{(\ell)}$. After solving these operators, we will use Eqs.~\eqref{eq:GLE} and \eqref{eq:dGLE} to solve for the orthogonal dynamics $\bfvF$ (continuous-time) and $\bfvW_k$ (discrete-time). 
Specifically, for the continuous-time system, we first set $t=0$ in Eq.~\eqref{eq:CE}, leading to 
\begin{equation}
    \dot{\bfvC}(0) = \bfvM \cdot \bfvC(0).
\end{equation}
Thus, we can solve for the Markov matrix $\bfvM= \dot{\bfvC}(0)\cdot \bfvC^{-1}(0)$. The estimation of $\dot{\bfvC}(0)$ can be made by finite-difference method, or alternatively, directly computed from the right-hand-side of the dynamical equation \eqref{eq:phiEvo} in the microscopic simulator. Next, we differentiate Eq.~\eqref{eq:CE} with respect to time $t$ once and obtain
\begin{equation}
    \ddot{\bfvC}(0) = \bfvK(0) \cdot \bfvC(0).
\end{equation}
The memory kernel evaluated at $t=0$ is $\bfvK(0)= \ddot{\bfvC}(0)\cdot \bfvC^{-1}(0)$. Again, $\ddot{\bfvC}(0)$ can either be accessed directly in the simulator, or estimated by finite-difference data. Then, we set $t=\delta \ll 1$ to Eq.~\eqref{eq:CE} and approximate the memory integration by trapezoidal rule  
\begin{equation}
    \int_0^\delta \bfvK(\delta-s) \cdot \bfvC(s) \dd s \approx \frac{\delta}{2} \l(\bfvK(0) \cdot \bfvC(\delta) + \bfvK(\delta) \cdot \bfvC(0)\r)
\end{equation}
to solve for $\bfvK(\delta) = \l[2 \dot{C}(\delta)- 2 \bfvM\cdot \bfvC(\delta) - \delta \bfvK(0)\cdot \bfvC(\delta)\r] \cdot \bfvC^{-1}\l(0\r)$. Similarly, we can recursively solve for $\bfvK(\l(k+1\r)\delta)$, $k \in \mathbb{N}$, as functions of previously obtained $\bfvK(\l(\ell \r)\delta)$, $\ell\le k$, and the measured correlation matricies $\bfvC$:
\begin{align}
{}& \quad \bfvK((k+1) \delta) =\\{}& 2 \l[ \frac{\dot{\bfvC}(k\delta) -  \bfvM \cdot \bfvC\l(k \delta\r)}{\delta} + \sum_{\ell=1}^{k} \bfvK\l(\ell \delta \r) \cdot \bfvC\l(\l(k-\ell\r) \delta \r) + \frac{ \bfvK\l(0\r)\cdot \bfvC\l(k\delta\r)}{2} \r] \cdot \bfvC^{-1}\l(0\r). \nonumber
\end{align}
Once $\bfvM$ and $\bfvK$ are solved, we use Eq.~\eqref{eq:GLE} and the measured trajectory to solve for the orthogonal dynamics $\bfvF(t)$.  A detailed description of our proposed procedure is presented as Algorithm \ref{alg:1}.

\begin{algorithm}[ht]
\caption{Data-driven learning of the continuous-time Mori--Zwanzig operators. This algorithm uses a long trajectory of the reduced-order dynamics to estimate the continuous-time Markov transition matrix $\bfvM$, memory kernel $\bfvK(s)$, and orthogonal dynamics $\bfvF(t)$ in the Generalized Langevin Equation \eqref{eq:GLE}. The algorithm requires the trajectories of $M$ \emph{a priori} selected observables $\l\{g_i\r\}_{i=1}^M$, measured at finely and evenly discretized times $t=k\delta$, $\delta \ll 1$, $k=0\ldots N-1$ along the long ($N\gg 1$) trajectory. A number $0 \le h < N$ is required to specify the longest estimated horizon of the memory kernel $\bfvK$. The algorithm will deliver estimates of the (1) Markov transition matrix $\bfvM$, (2) memory kernel at the discretized time points $\bfvK(k\delta)$, and (3) the orthogonal dynamics $\bfvF(k\delta\vert i)$, evaluated at discrete times $k=0\ldots h-1$, conditioned on the system started at the $i^\text{th}$ snapshot (i.e., the system's initial condition was set at $\bfvg\l(i\delta\r)$).}
\label{alg:1}
\begin{algorithmic}
\For{$k$ in $\l\{0,\ldots h+1\r\}$}
\State{$C_{ij}\l(k \delta\r)\leftarrow \frac{1}{N-k}\l[\sum_{\ell=0}^{N-k-1} g_i( \l(k+\ell\r) \delta)\times g_j(  \ell \delta)\r] $}
\EndFor
\For{$k$ in $\l\{1,2\r\}$}
\State{$C_{ij}\l(-k \delta\r)\leftarrow C_{ji}\l( k \delta\r)$}
\EndFor
\For{$k$ in $\l\{-1,0,\ldots h\r\}$}
\State{$\dot{C}_{ij}\l(k \delta\r)\leftarrow \frac{1}{2\delta}\l[C_{ij}\l((k+1)\delta\r)-C_{ij}\l((k-1)\delta\r)\r]$} 
\EndFor
\For{$k$ in $\l\{0,\ldots h\r\}$}
\State{$\ddot{C}_{ij}\l(0\r)\leftarrow \frac{1}{2\delta}\l[\dot{C}_{ij}\l(\delta\r)-\dot{C}_{ij}\l(-\delta\r)\r]$} 
\EndFor
\State{$\bfvM\leftarrow \dot{\bfvC}(0) \cdot \bfvC^{-1}\l(0\r) $} 
\State{$\bfvK\l(0\r)\leftarrow \ddot{\bfvC}(0) \cdot \bfvC^{-1}\l(0\r)$} 
\For{$k$ in $\l\{1,\ldots h\r\}$}
\State{$\bfvK(k \delta) \leftarrow 2 \l[ \frac{\dot{\bfvC}(0) -  \bfvM \cdot \bfvC\l(k \delta\r)}{\delta} + \sum_{\ell=1}^{k-1} \bfvK\l(\ell \delta \r) \cdot \bfvC\l(\l(k-\ell\r) \delta \r) + \frac{ \bfvK\l(0\r)\cdot \bfvC\l(k\delta\r)}{2} \r] \cdot \bfvC^{-1}\l(0\r) $} 
\EndFor
\For{$i$ in $\l\{0,\ldots N-h-1\r\}$}
\For{$k$ in $\l\{0,\ldots h\r\}$}
\State{$m \leftarrow \frac{\delta}{2} \l[ \bfvK\l(0\r)\cdot \bfvg\l(k\delta \r) + 2 \sum_{\ell=1}^{k-1} \bfvK\l(\ell \delta \r) \cdot \bfvC\l(\l(k-\ell\r) \delta \r) + \bfvK\l(k\delta \r)\cdot \bfvg\l(0 \r)\r]$} 
\State{$\bfvF(k\delta\vert i) \leftarrow \frac{1}{\delta}\l[\bfvg\l(\l(k+1\r)\delta \r) -\bfvg\l(k\delta \r) \r] -\bfvM\cdot\bfvg\l(k\delta \r) + m $}
\EndFor
\EndFor
\Return{$\bfvM$, $\bfvK(k\delta)$, $\bfvF(\ell\delta\vert i)$}
\end{algorithmic}
\end{algorithm}

As for the discrete-time dynamics, we use Eq.~\eqref{eq:dCE}. Setting $k=0$ in \eqref{eq:dCE}, $\bfvC\l(\Delta\r)=\bfvO_\Delta^{(0)} \bfvC(0)$ indicates that $\bfvO_\Delta^{(0)}=\bfvC\l(\Delta\r)\cdot  \bfvC^{-1}(0)$, exactly the approximate Koopman operator that one would obtain by carrying out EDMD analysis (cf.~Sec.~\ref{sec:koopman}). Then, recursively, we can solve $\bfvO_\Delta^{(k)}$, $k\in\mathbb{N}$ in terms of the correlation matrices $\bfvC$ and previously solved lower-order $\Omega^{(\ell)}_\Delta$, $\ell < k$, using Eq.~\eqref{eq:dCE}: 
\begin{equation}
\bfvO_\Delta^{(k)}=\l(\bfvC\l(\l(k+1\r)\Delta \r)-\sum_{\ell=0}^{k-1} \bfvO^{(\ell)}_\Delta \cdot \bfvC\l(\l(k-\ell\r)\Delta \r) \r)\cdot \bfvC^{-1}(0).
\end{equation} 
Once the operators $\bfvO^{(k)}_\Delta$'s are solved, we use Eq.~\eqref{eq:dGLE} and the measured trajectory to solve for the discrete-time noise $\bfvW$. A detailed description of the procedure is presented as Algorithm \ref{alg:2}.

\begin{algorithm}[ht]
\caption{Data-driven learning of the discrete-time Mori--Zwanzig operators. This algorithm estimates the discrete-time operators $\bfvO_\Delta^{\l(\ell\r)}$ and the orthogonal dynamics $\bfvW$ in the discrete-time Generalized Langevin Equation \eqref{eq:dGLE} from a long trajectory of the dynamical simulations of the dynamical system. The algorithm requires the snapshots of a set of $M$ \emph{a priori} selected observables $\l\{g_i\r\}_{i=1}^M$ measured at evenly distributed times $t=k\Delta$, $k=0\ldots N-1$,  along a long ($N\gg 1$) trajectory. In contrast to Algorithm \ref{alg:1}, the time separation of the snapshots $\Delta$ does not necessarily to be small. A number $0\le h < N$ is required to specified the highest order of the discrete-time operators ($\bfvO^{\l(h\r)}$). The algorithm delivers estimates of (1) the discrete-time operators $\bfvO(k\Delta)$ and (2) the orthogonal dynamics $\bfvW\l(k \Delta \vert i \Delta\r)$, $k=0,1,\ldots h$ conditioned on the system started at the $i^\text{th}$ snapshot (i.e., the system's initial condition was set as $\bfvg\l(i \Delta\r)$).}
\label{alg:2}
\begin{algorithmic}
\For{$k$ in $\l\{0,\ldots h+1\r\}$}
\State{$C_{ij}\l(k \Delta\r)\leftarrow \frac{1}{N-k} \l[\sum_{\ell=0}^{N-k-1} g_i( \l(k+\ell\r) \Delta )\times g_j( \ell \Delta)\r]$} 
\EndFor
\State{$\bfvO_\Delta^{(0)}\leftarrow \bfvC(\Delta) \cdot \bfvC^{-1}\l(0\r) $}  
\For{$k$ in $\l\{1,\ldots h\r\}$} 
\State{$\bfvO_\Delta^{\l(k\r)} \leftarrow \left [\bfvC\l(\l(k+1\r)\Delta\r) - \sum_{\ell=0}^{k-1} \bfvO^{\l(\ell\r)} \cdot \bfvC \l(\l(k-\ell\r) \Delta \r) \right] \cdot \bfvC^{-1}(0)$} 
\EndFor
\For{$i$ in $\l\{0,\ldots N-h-1\r\}$} 
\For{$k$ in $\l\{0,\ldots h\r\}$} 
\State{$\bfvW(k\Delta\vert i\Delta) \leftarrow \bfvg\l( i+k+1 \r) - \sum_{\ell=0}^{k} \bfvO^{\l(\ell\r)} \cdot \bfvg \l(\l(i+k-\ell\r) \Delta \r) $}  
\EndFor
\EndFor
\Return{$\bfvO_\Delta^{(k)}$, $\bfvW(k\Delta\vert i\Delta)$}
\end{algorithmic}
\end{algorithm}

\begin{figure}[ht]
\centering
\includegraphics[width=0.96\textwidth]{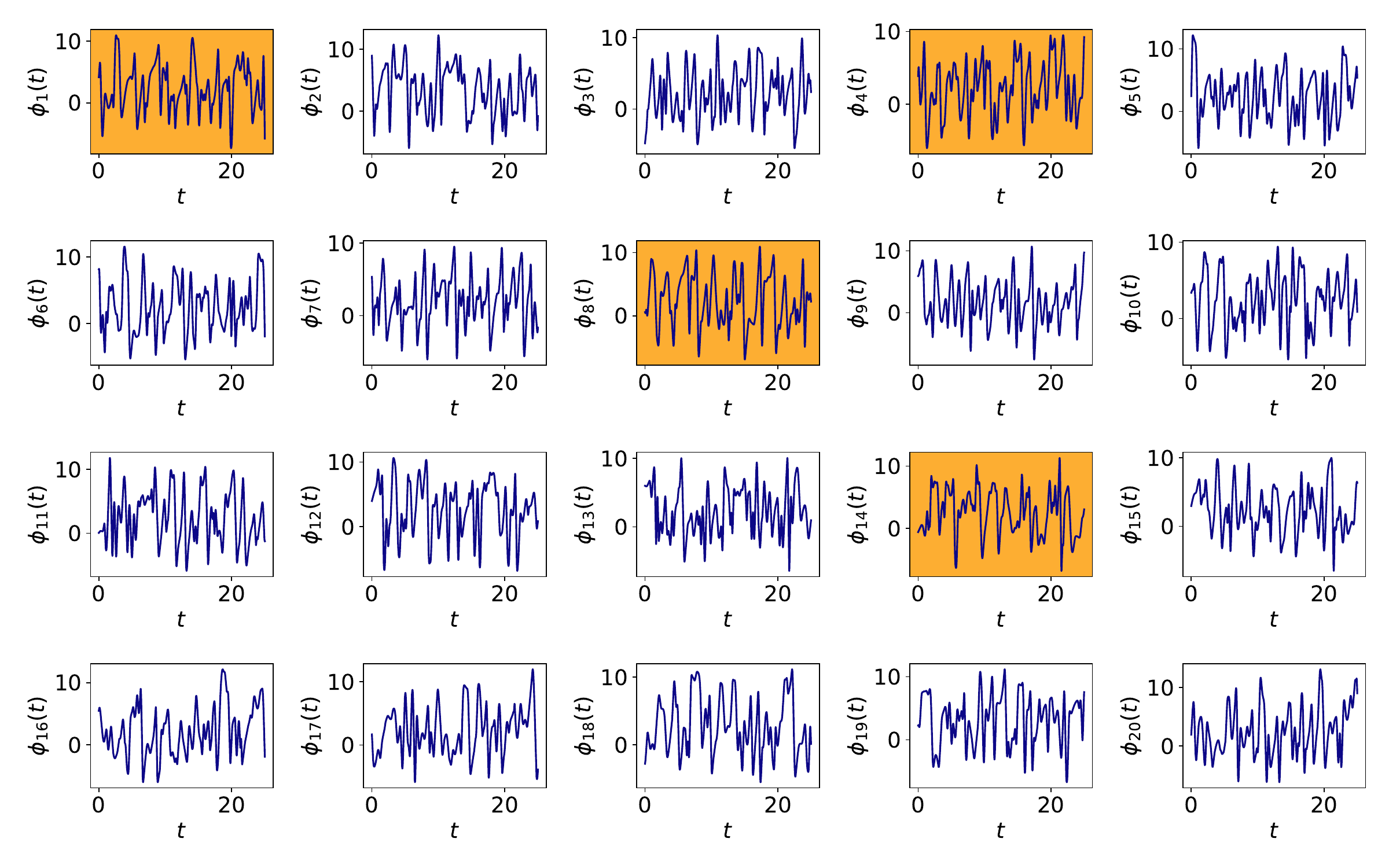}
\caption{The dynamics of the physical-space variables $\l\{\phi_i\r\}_{i=1}^{20}$ of the Lorenz '96 system \eqref{eq:Lorenz96}. The highlighted panels are those chosen variables which serve as the observables $g_1:=\phi_1$, $g_2:= \phi_4$, $g_3:=\phi_8$, and $g_4:=\phi_{14}$. Because the observables are not zero-meaned, we also include a constant function $g_0:=1$. For the reference of the reader, the estimated Lyapunov time $\approx 0.622$. } \label{fig:LorenzDynamics}
\end{figure}

\begin{figure}[ht]
\centering
\includegraphics[width=0.96\textwidth]{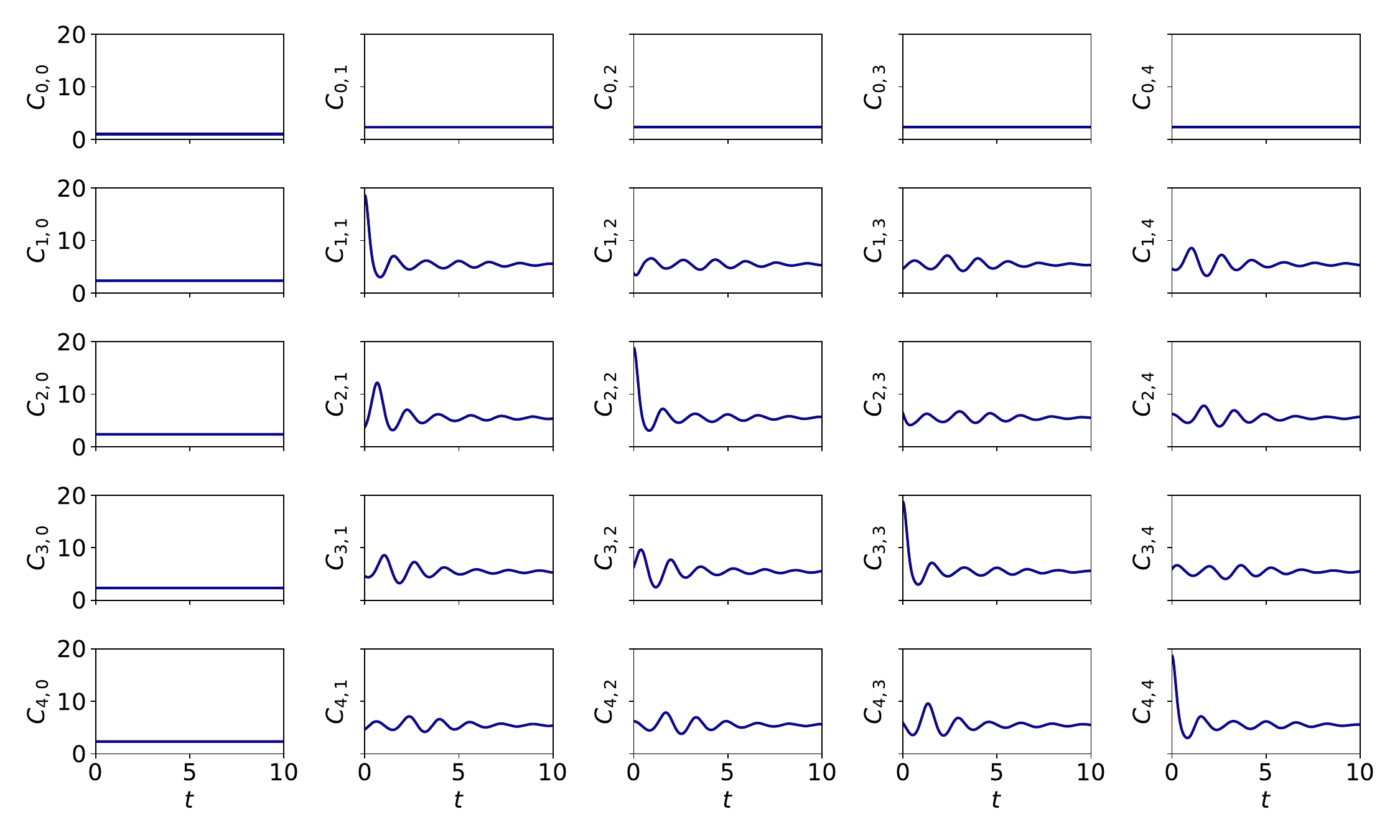}
 \caption{The two-time correlation function $\bfvC(t)$ computed from snapshots along a long ($10^5$) trajectory $C_{ij}(k\delta )=10^{-5} \times \sum_{\ell=1}^{10^5-k}
 g_i( \l(k+\ell\r) \delta )\, g_j( k\delta )$.  The estimated Lyapunov time $\approx 0.622$.} \label{fig:correlationFunctions}
\end{figure}

\section{Numerical experiment} \label{sec:experiments}
\subsection{Test model} \label{sec:example} 
In this section, we present the application of our proposed algorithms. Our aim is to illustrate that the usage and the capability of the algorithms to extract the Mori--Zwanzig operators from a long trajectory of simulation.

Throughout this section, we adopt the Lorenz '96 system \cite{lorenz1996predictability} as the test problem. The system consists of $N$ physical-space variables which evolve nonlinearly by
\begin{equation}
\frac{\text{d}}{\text{d}t} \phi_i \left(t\right) = \left(\phi_{i+1}-\phi_{i-2}\right) \phi_{i-1} - \phi_i + F, \ i=1\ldots N, \label{eq:Lorenz96}
\end{equation}
with the periodic boundary condition $\phi_{-1} = \phi_{N-1}$, $\phi_{0} = \phi_{N}$, and $\phi_{N+1} = \phi_{1}$. We fixed the model parameter $N=20$ and $F=8$, with which the system has a chaotic behavior. 

We define our observables to be $g_1(t):=\phi_1(t)$, $g_2(t):= \phi_4(t)$, $g_3(t):=\phi_8(t)$, and $g_4(t)=\phi_{14}(t)$. We also include a constant function $g_0 =1$ because the long-time average of the observables are not zero-meaned. We remark that the results below are conditioned on the choice of the observables we choose. We use the general-purpose integrator LSODA (by \texttt{scipy.integrate.solve\_ivp}) to solve the evolutionary equation \eqref{eq:Lorenz96}. LSODA uses adaptive time steps for controlling the error of the numerical integration, and can handle stiff ODE systems. We chose a randomized initial condition, integrated the trajectory until $t=10^5$, and recorded the snapshot of the observables every $\delta=0.01$. The total length $t=10^5$ was deemed sufficient from the convergence of the computed correlation matrix $\bfvC$ between these chosen observables. We checked that the choice of the initial condition does not affect the obtained two-time correlations. Numerically, the trajectory is long enough such that the contribution of the initial transient behavior of the trajectory converging to the attractor is negligible. A short-time behavior of the dynamical system is shown in Fig.~\ref{fig:LorenzDynamics}, where the observables are highlighted. We numerically computed the Lyapunov exponents of the system and determined that the largest Lyapunov exponent $\approx 1.608 $, which corresponds to a Lyapunov time $\approx 1/1.608 \approx 0.622$. The estimated Kaplan--Yorke dimension \cite{frederickson1983LyapunovDimensionStrange} of the system is $\approx 13.37$. 

\begin{figure}[ht]
\centering
\includegraphics[width=0.96\textwidth]{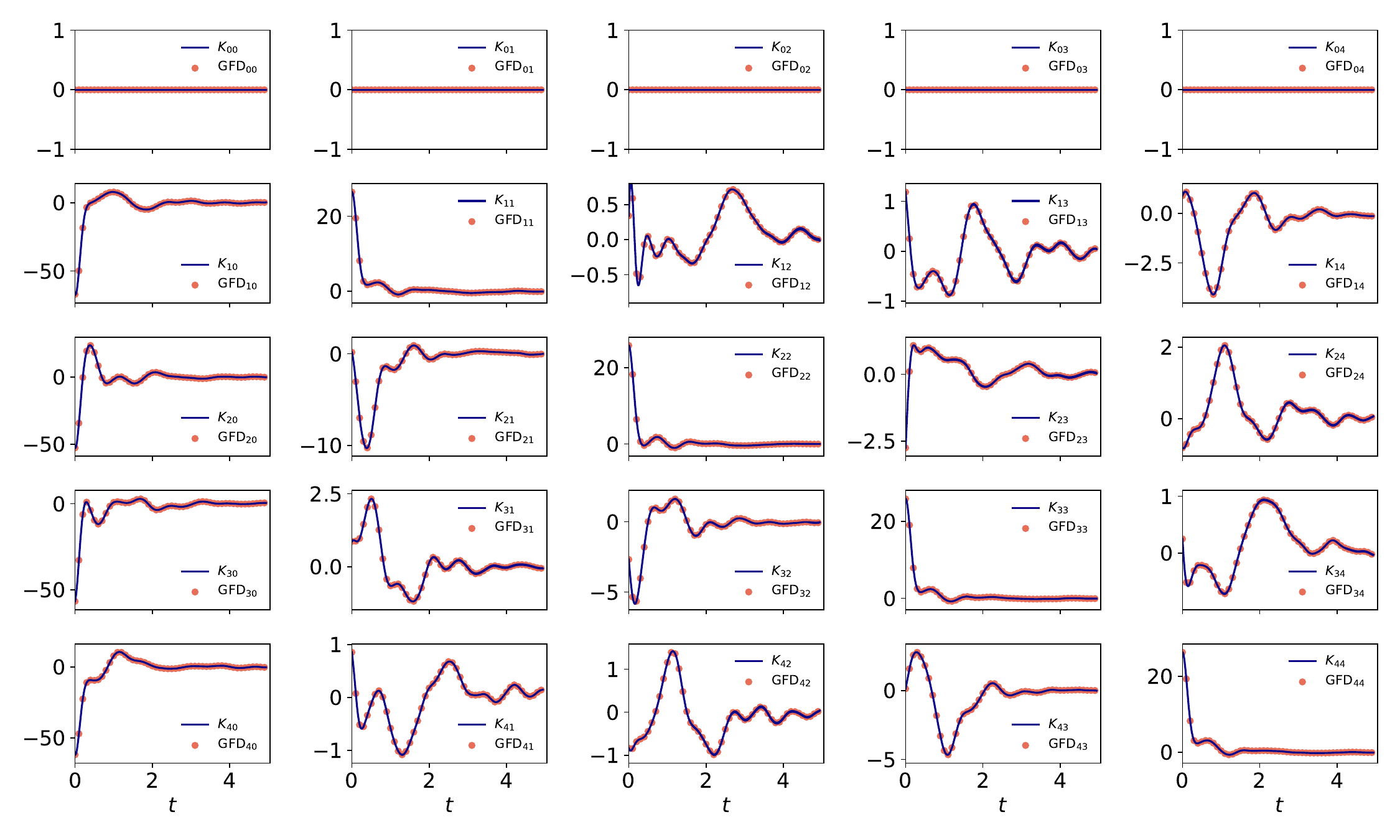}
\caption{The memory kernel $\bfvK(t)$ is illustrated as the solid line. To verify the Generalized Fluctuation-Dissipation (GFD) relationship, the right-hand-side of Eq.~\eqref{eq:GFD}, $\text{GFD}_{ij}:=\l[\l\langle \bfvF\l(t\r) \cdot \bfvF^T\l(0\r) \r\rangle\cdot \bfvC^{-1}\l(0\r)\r]_{ij}$, is calculated and plotted as the discrete dots, which perfectly align with the memory kernel $\bfvK$. The timescale of nontrivial memory kernel exceeds the estimated Lyapunov time $\approx 0.622$.} \label{fig:K-and-FD}
\end{figure}

We use the collected snapshots of the observables to compute two-time correlation function up to a lag $t=10$, as shown in Fig.~\ref{fig:correlationFunctions}. Then, we apply Algorithm \ref{alg:1} to the the numerically computed $\bfvC\l(t\r)$ to extract the continuous-time Markov matrix 
\begin{equation}
\bfvM = \begin{bmatrix}
0.0 & 0.0 & 0.0 & 0.0 & 0.0 \\
1.0 & -0.052 & -0.416 & 0.215 & -0.175 \\
0.652 & 0.372 & 0.099 & -0.795 & 0.047 \\
-2.333 & -0.066 & 0.787 & -0.068 & 0.342 \\
0.338 & 0.171 & 0.043 & -0.378 & 0.021 
\end{bmatrix},\label{eq:Markov}
\end{equation}
and the memory kernel $\bfvK(t)$, illustrated in Fig.~\ref{fig:K-and-FD}. With the the calculated kernel $\bfvK$, Algorithm \ref{alg:1} also quantifies the orthogonal dynamics, $\bfvF\l(t, \bfvPhi\l(s\r)\r)$ for $t\ge s$, which allows us to calculate the right-hand side of the Generalized Fluctuation-Dissipation relationship \eqref{eq:GFD}. The GFD, which servers as a stringent self-consistent condition, is numerically verified and presented in Fig.~\ref{fig:K-and-FD}.

\begin{figure}[ht]
\centering
\includegraphics[width=0.96\textwidth]{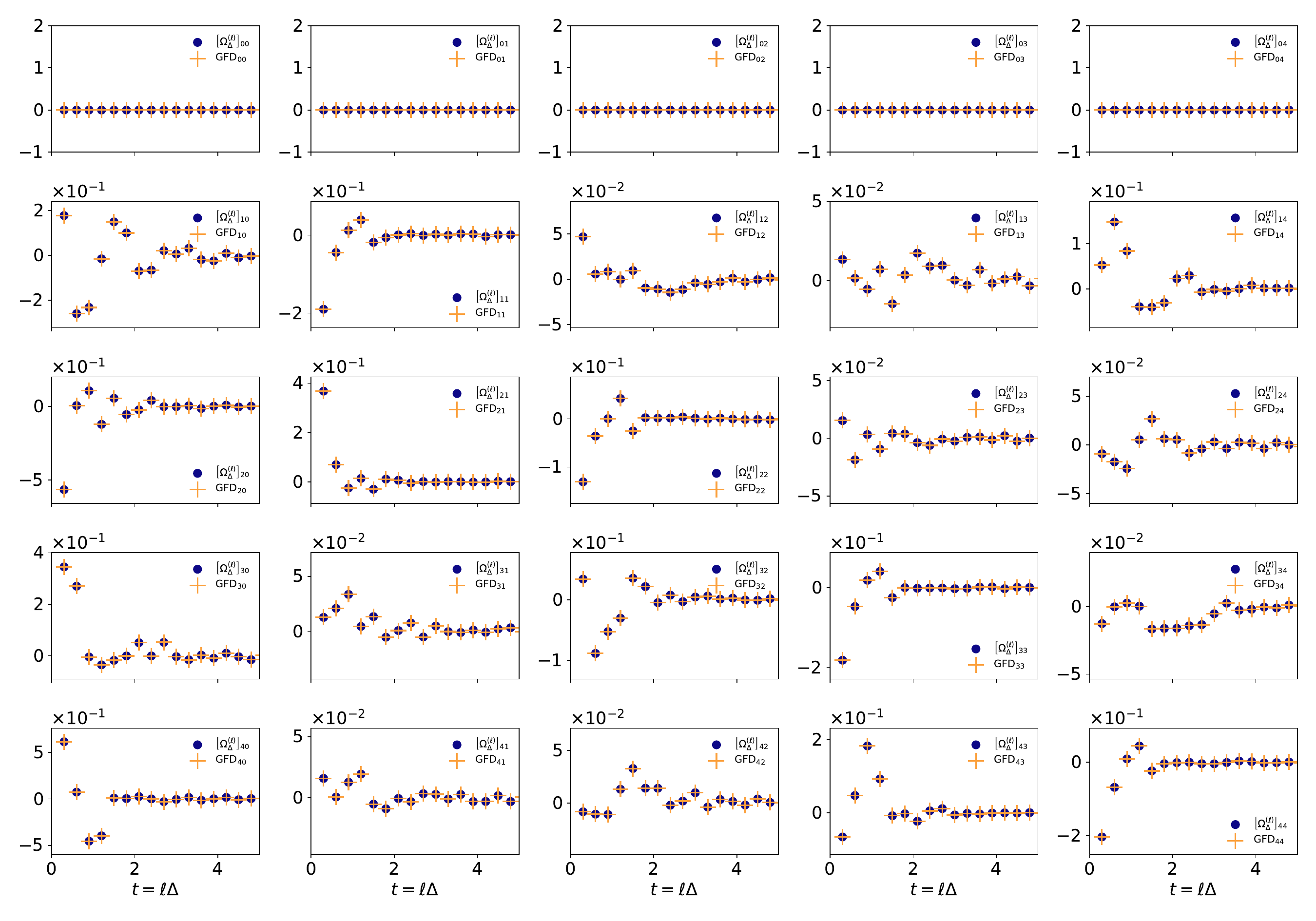}
\caption{We present the extracted higher-order discrete-time operators $\bfvO_\Delta^{(\ell)}$, $\ell \ge 1$ in discrete dots. To verify the discrete-time Generalized Fluctuation-Dissipation (GFD) relationship, the right-hand-side of Eq.~\eqref{eq:discrete-time-GFD1}, $\text{GFD}_{ij}:= \l[ \l\langle \bfvW_\ell \cdot \bfvW_0^T \r\rangle \cdot \bfvC^{-1}\l(-\Delta\r)\r]_{ij}$, is calculated and plotted as the discrete pluses, which align with the memory kernel $\bfvO^{(\ell)}_\Delta$. } \label{fig:Omega-and-FD}
\end{figure}

We also applied the Algorithm \ref{alg:2} to extract the discrete-time operators $\bfvO_\Delta^{(\ell)}$. We first fix $\Delta = 30 \delta$ and identify the lowest order $\bfvO^{(0)}_\Delta$ which serves like the Markov transition matrix in the discrete-time formulation:
\begin{equation}
\bfvO_\Delta^{(0)} = \begin{bmatrix}
1.0 & 0.0 & 0.0 & 0.0 & 0.0 \\
1.827 & 0.294 & -0.062 & 0.036 & -0.053 \\
1.378 & 0.173 & 0.332 & -0.108 & 0.016 \\
0.925 & -0.015 & 0.27 & 0.284 & 0.064 \\
1.771 & 0.026 & 0.012 & -0.105 & 0.311 \\
\end{bmatrix}, \label{eq:dMarkov}
\end{equation}
We present the higher-order operators, $\bfvO_{\Delta}^{(\ell)}$ in Fig.~\ref{fig:Omega-and-FD}. Algorithm \ref{alg:2} also quantifies the discrete-time orthogonal dynamics $\bfvW_k$, which we used to evaluate the right-hand side of the discrete-time GFD relationship \eqref{eq:discrete-time-GFD1}. Again, the stringent self-consistent GFD illustrates that our numerical analysis quantifies the operators accurately.

For a comparison between the continuous-time and discrete-time operators, we compute the propagator $\exp(\Delta\times \bfvM)$ using the continuous-time Markov operator $\bfvM$ computed in \ref{eq:Markov}:
\begin{equation}
\exp\l(\Delta \times \bfvM\r)  = \begin{bmatrix}
1.0 & 0.0 & 0.0 & 0.0 & 0.0 \\
0.256 & 0.976 & -0.117 & 0.08 & -0.049 \\
0.296 & 0.114 & 0.995 & -0.234 & -0.001 \\
-0.658 & -0.003 & 0.236 & 0.946 & 0.103 \\
0.149 & 0.052 & -0.004 & -0.111 & 0.999 \\
\end{bmatrix}. \label{eq:Markov30}
\end{equation}
 The difference between the lowest-order Markov operators, Eqs.~\eqref{eq:dMarkov} and \eqref{eq:Markov30}, illustrate the difference between the continuous-time and the discrete-time formulation. The discrete Markov operator $\bfvO^{(0)}_\Delta$ is identical to the approximate Koopman operator if one would carry out the EDMD \cite{williams2015data} with a time separation $\Delta=30\delta=0.3$. Nevertheless, as pointed out in Sec.~\ref{sec:koopman}, although the operator $\bfvO^{(0)}_\Delta$ can always be estimated by data (using Algorithm \ref{alg:2}) and is optimal to predict one-step $\Delta$ into the future, it cannot be approximated by exponentiating the instantaneous Markov operator of the continuous-time formulation, $\exp\l(\Delta\times \bfvM\r)$. Our analysis in Appendix \ref{app:discrete-time-1} shows that $\bfvO^{(0)}_\Delta$ contains not only the effect of continuous-time Markov operator $\bfvM$ which is the exact Koopman operator of the dynamics, but also the effect of the continuous-time memory kernel $\bfvK(s)$, $0\le s \le \Delta$.

The continuous-time and discrete-time memory kernels shown in Figs.~\ref{fig:K-and-FD} and \ref{fig:Omega-and-FD} show similar behavior. Note that in the GLE \eqref{eq:GLE}, it is conventional that the memory kernel $\bfvK$ comes with an overall negative sign in front, but in the discrete formulation (Eq.~\eqref{eq:dGLE}), it is more natural not to put the $-1$ only in front of the $\ell\ge 1$ terms. Thus, to make comparison between Figs.~\ref{fig:K-and-FD} and \ref{fig:Omega-and-FD}, one of them shall be flipped upside-down. We point out a few important observations in these numerical estimations of the memory kernels. First, it shows that in both formulations, the memory kernels decay to zero at a finite timescale. The finite timescale of the memory kernel indicates that operationally, we do not necessarily need the full history of the system from $t=0$ until the current time to make prediction---the trajectory with the finite timescale is sufficient. Secondly, the analysis shows that the memory kernels could behave very non-trivially, and the kernels are not likely captured by simple models. Thirdly, the discrete-time memory kernel seems to be a coarse-grained and smoothed-out picture of the continuous-time kernel. 

\begin{figure}[ht]
\includegraphics[width=0.96\textwidth]{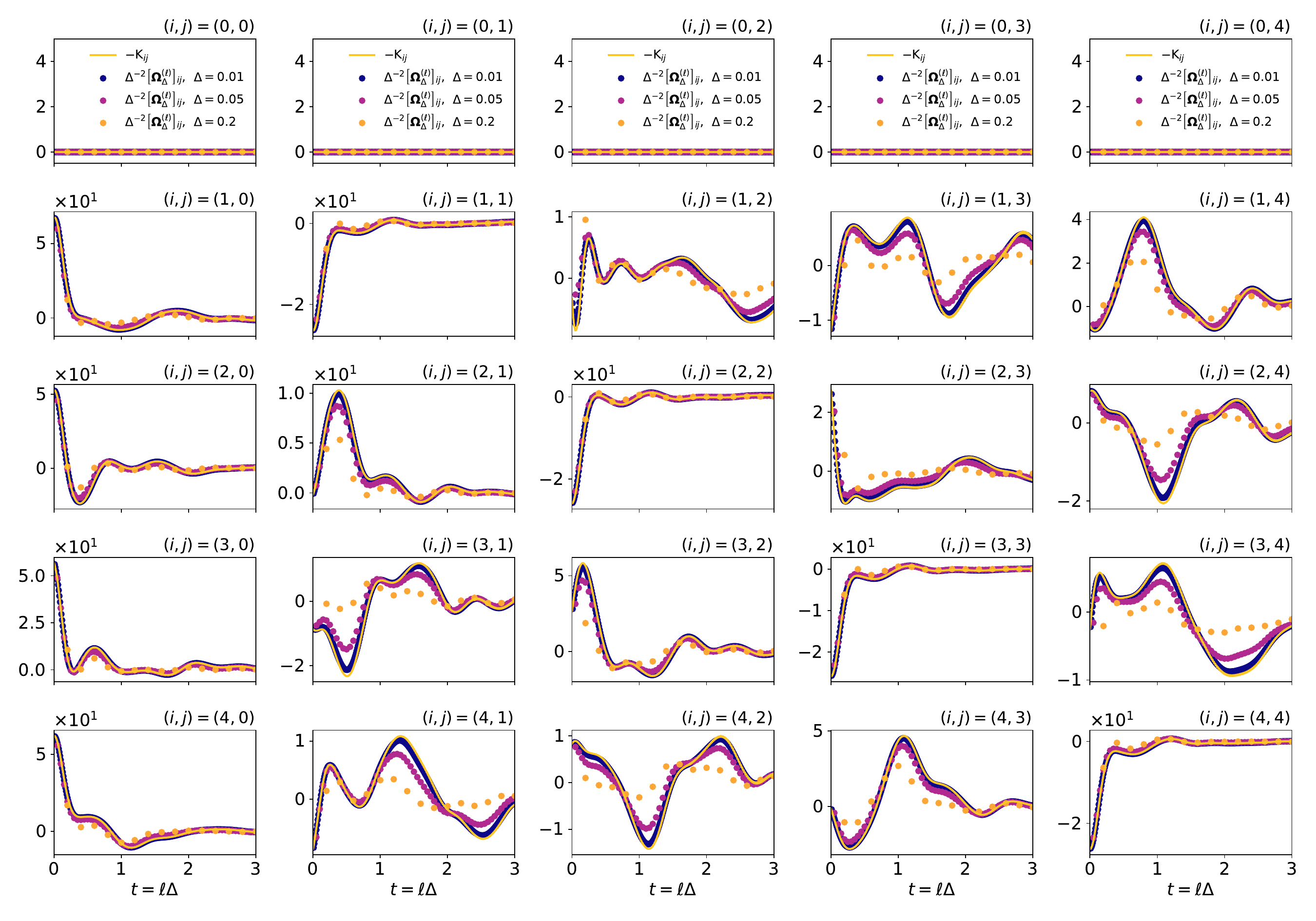}
\caption{Comparisons between the discrete-time memory operators $\bfvO_\Delta^{(\ell)}$, $\ell \ge 1$ with various $\Delta$ and the continuous-time memory kernel $-\bfvK(t)$. Note that the continuous-time memory kernel in the generalized Langevin equation \eqref{eq:GLE} has a sign difference to the discrete-time formulation \eqref{eq:dGLE}. To make comparable visualization, we compare the discrete-time operators to the negative of continuous-time memory kernel, $-\bfvK$.} \label{fig:Omegas-and-K}
\end{figure}

To investigate further into the $\Delta$-dependence smoothing of the discrete memory operators, we calculate $\bfvO_\Delta^{(\ell)}$, $\ell \ge 1$ with $\Delta=0.01$, $0.1$, and $0.2$. Note that the smallest $\Delta=0.01$ corresponds to the smallest time resolution $\delta=0.01$, which we used to compute the continuous-time operators. To properly scale and compare their behavior, we need to scale the discrete operators by $\Delta^
{-2}$. The first scaling $\Delta$ comes from the fact that for a fixed physical memory timescale, a larger $\Delta$ comes with fewer snapshots in the discrete sum of the past history. Another way to understand this scaling is that $\bfvO$ in Eq.~\eqref{eq:dGLE} is analogous to $-\bfvK(s) \dd s$ in Eq.~\eqref{eq:GLE}, and the scaling comes from $\Delta \sim \dd s$. The second scaling comes from the fact that the discrete operators map the system forward to $\Delta$. We present the scaled operators, $\Delta^{-2} \bfvO^{(\ell)}_\Delta$ on the same figure \ref{fig:Omegas-and-K}, which shows that when $\Delta=\delta=0.01$, the calculated discrete-time operator by Algorithm \ref{alg:2} converges to the continuous-time kernel calculated using continuous-time Algorithm \ref{alg:1}. The smoothing as we increase $\Delta$ can also be observed. With this comparison, we recommend the discrete-time formulation as it requires less inputs but achieves comparable results of the continuous-time formulation which needs estimation of $\dot{\bfvC}(0)$ and $\ddot{\bfvC}(0)$, when the time separation $\Delta$ is set equal to the fine discretization $\delta$ of the continuous-time model. 

\subsection{Advantage of the Mori--Zwanzig formalism in prediction}\label{sec:errorAnalysis}

In this section, we compare different numerical procedures for making prediction to illustrate the advantage of the Mori--Zwanzig formalism over plain Koopman analysis when the set of basis functions is not complete. Specifically, we consider the following problem setup. Suppose we obtain snapshots of a set of observables along a long trajectory. These snapshots, separated by a fine time resolution $\delta$, were used to compute either the approximate Koopman operator (cf.~Sec.~\ref{sec:koopman}) or the Mori--Zwanzig operators (cf.~Algorithm \ref{alg:2}). Next, suppose we were given the snapshots of the observables along a segment of trajectory of length $\zeta=m\delta$, $m\in \mathbb{N}$. We denote the snapshots by $\l\{\bfvg(k\delta)\r\}_{k=1}^m$, noting that $\bfvg$ is an $M\times 1$ column vector. We are interested in using the given snapshots to predict the observables $\eta=n\delta$, $n\in \mathbb{N}$ in the future. Specifically, we are interested in comparing the prediction errors of Koopman and discrete-time Mori--Zwanzig formulations:\\
\noindent {\bf 1.~The recursive Koopman computation:} With this approach, we use the EDMD \cite{williams2015data} to compute the approximate Koopman operator $\bfvK_{\Delta }^{\rm Koop}$ with a time separation $ \Delta = k \delta$. Because the aim is to predict $n\delta$ into the future, $k$ is restricted to the set of divisors of $n$. When $k=1$, the learned approximate Koopman operator would be closest to the true Markov operator in the Mori--Zwanzig formalism. When $k=n$, the time separation is chosen to be exactly the predictive horizon. We then apply $n/k$ times the learned Koopman operator to advance the last known observables to make prediction, i.e., $\bfvg\l(\l(m+n\r)\delta\r) \approx \l(\bfvK_{\Delta}^{\rm Koop}\r)^{n/k} \cdot \bfvg\l(m\delta\r)$. \\
\noindent{\bf 2.~The recursive discrete-time Mori--Zwanzig computation:} With this approach, we use Algorithm \ref{alg:2} to compute the Mori--Zwanzig operators $\bfvO^{(\ell)}_\Delta$. We are interested in different magnitudes of $\Delta$, noting a restriction that they must be multiples of the finest time resolution $\delta$. We chose $\Delta = 0.01$, $0.02$, $0.05$, $0.1$, and $0.2$. We are also interested in the predictive error as a function of the memory length, which is restricted as multiples of $\Delta$. We will integrate Eq.~\eqref{eq:dGLE} to advance $\bfvg(t)$ to $\bfvg(t+\Delta)$, noting that it is not possible to estimate the exact noise $\bfvW_k$. Instead of injecting artificially generated samples from a noise model, we set $\bfvW_k$'s to zero. Note that the steps needed for such integration is also $\Delta$-dependent. For example, when $\Delta$ matches the predictive horizon ($\Delta=n\delta$), we only need to integrate Eq.~\eqref{eq:dGLE} once; when $\Delta$ is the finest time resolution ($\Delta=\delta$), we need to interatively integrate Eq.~\eqref{eq:dGLE} $n$ times and accumulate the prediction $\bfvg\l(\l(m+k\r)\delta\r)$, $1\le k < n$ as past history until the horizon is met.

We remark that when the memory length is set to zero in the recursive discrete-time Mori--Zwanzig approach, the method converges to the recursive Koopman approach because $\bfvK^{\rm Koop}_\Delta\equiv \bfvO^{(0)}_\Delta$. Thus, the second family of numerical procedures (with different settings of $\Delta$ and memory length) is a superset of the first one. 

We adopt the $L^2$-norm as the measure to compare errors between different methods. That is, suppose a method made a prediction $\bfvg^\text{pred}$ and suppose the ground truth is $\bfvg^\text{GT}$, the error is computed by
\begin{equation}
    \varepsilon^2 :=\l\Vert \bfvg^\text{pred} - \bfvg^\text{GT}\r\Vert_2^2 \equiv \sum_{i=1}^M (g_i^\text{pred}-g_i^\text{GT})^2.
\end{equation}
To collect the statistics of the prediction error, we generate $2\times 10^4$ samples of segments, each of which contains $m=500$ snapshots (separated by $\delta=0.01$), sampled from a long ($t=10^5$) test trajectory. Because we are interested in out-of-sample prediction, the long test trajectory used to generate test samples is different from the one we used to compute the correlation matrix (and the operators). Both trajectories were generated by integrating the same evolutionary equation \eqref{eq:Lorenz96} with two randomized initial conditions; the procedure ensures a consistent inner-product space which is defined by the long-time statistics. 

\begin{figure}[ht]
\centering
\includegraphics[width=0.96\textwidth]{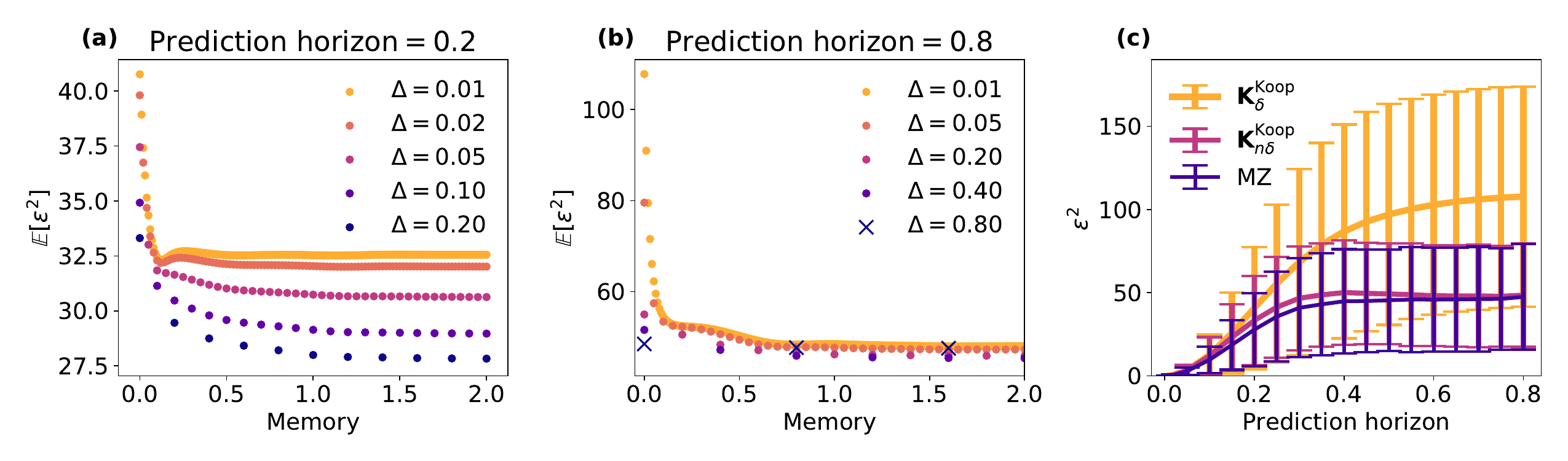}
\caption{Comparison of the prediction error of different numerical procedures. (a) The mean error of the Mori--Zwanzig formulation for predicting the observable $n\delta = 0.2$ into the future. (b) The mean error of the Mori--Zwanzig formulation for predicting the observable $n\delta = 0.8$ into the future. (c) We compare the statistics of the prediction error as the function of the prediction horizon $n \delta$ from (1) the approximate Koopman operator $\bfvK_{\delta}$, (2) the approximate Koopman operator $\bfvK_{n \delta}$, and (3) recursive Mori--Zwanzig computation. For the reader's reference, the estimated Lyapunov time $\approx 0.622$.} \label{fig:errorAnalysis}
\end{figure}

We present the error statistics from the Mori--Zwanzig with different memory and prediction horizon in Fig.~\ref{fig:errorAnalysis}. In Fig.~\ref{fig:errorAnalysis} (a) and (b), we present the average error over $2\times 10^4$ samples when the prediction horizon is chosen as $n\delta=0.2$ and $0.8$ ($20$ and $80$ finest time steps, $\delta=0.01$), respectively. The recursive Koopman computation corresponds to those points with the zero memory length. A few important observations can be made. First, for approximate Koopman computation, the most accurate way is to match the time separation $\Delta$ to the prediction horizon $n\delta$. This observation is consistent to our reasoning in Sec.~\ref{sec:CE+OP}, that the optimal prediction $n\delta$ into the future is $\bfvC(n\delta) \cdot \bfvC^{-1}(0)$, which is the approximate Koopman operator $\bfvK^\text{Koop}_{n\delta}$ (cf.~Sec.~\ref{sec:koopman}). Secondly, as the length of the past history is increased, the error decreases but levels off at a finite timescale which depends on the choice of $\Delta$. The improvement comes from an important fact that the given segment of trajectory $\bfvg(k\delta)$, $k=1\ldots m$, contains information in both the parallel and orthogonal spaces; see decomposition Eqs.~\eqref{eq:PE} and \eqref{eq:OE}. By integrating GLE forward in time with the segment of trajectory, we can evolve both the parallel and orthogonal dynamics beyond $t=m\delta$, despite the fact that we cannot access the unknown orthogonal dynamics $\bfvF(t\ge m\delta)$, which is thus set to be zero operationally for making predictions. In contrast, the Koopman approach always predicts from the last snapshot at $t=m\delta$ and can only optimally predict in the parallel space by Eq.~\eqref{eq:PE} and misses out more orthogonal contributions. Interestingly, the optimal choice of $\Delta$, the separation between snapshots for learning, depends nontrivially on the memory length and the prediction horizon in the Mori--Zwanzig formulation. For a prediction horizon $0.2$ (Fig.~\ref{fig:errorAnalysis} (a)), it is optimal to match $\Delta=0.2$ regardless of the memory length; however, for a prediction horizon $0.8$ (Fig.~\ref{fig:errorAnalysis} (b)), it is more advantageous to adapt a mismatched $\Delta=0.4$ in the long-memory kernel regime. Regardless, in all cases, we found that including the memory contribution improves the prediction, compared to the memory-less Koopman prediction. Finally, we observed that the memory length does not need to be long before the accuracy levels off. 

Consistent with these results, it is not expensive to improve the accuracy using the discrete-time Mori--Zwanzig formulation: we only need to store a finite length of snapshots to carry out the discrete sum in \eqref{eq:dGLE}. In Fig.~\ref{fig:errorAnalysis} (c), we present the error as a function of the prediction horizon $n\delta$. In addition to the average error, we also visualize the standard deviation of the $\varepsilon^2$ in $2\times 10^4$ sample segments. We compare three methods: Koopman with the smallest time separation $\Delta = \delta$, with the matched time separation $\Delta=n\delta$, and discrete Mori--Zwanzig with $\Delta=n\delta$ with a fixed memory length $0.24$. We conclude that, the Mori--Zwanzig formulation is consistently more accurate in making predictions into the future, and if one must carry out the approximate Koopman analysis, it is best to match the time separation $\Delta$ to that of the prediction horizon---learning an approximate Koopman operator with a short time separation $\Delta$ and recursively applying $\bfvK^\text{Koop}_{\Delta<n\delta}$ invokes more error. 

Finally, it is worth pointing out that similar to the Koopman analysis, the quality of the prediction crucially depends on the selection of the observables. In our test problems above, we only considered five simple observables $\l\{1, \phi_1, \phi_4, \phi_{8}, \phi_{14}\r\}$ and performed apples-to-apples comparison between the Koopman and Mori--Zwanzig predictions. We emphasize on the result that the Mori--Zwanzig, with the same functional basis, improves the prediction of the plain Koopman analysis, despite that the absolute error of the Mori--Zwanzig prediction is still large (see Fig.~\ref{fig:errorAnalysis} (a) and (b)). To further improve the prediction, one then needs to optimize a set of observables, which is an important procedure but not the focus of this article. 

\section{Discussion}\label{sec:discussion}

In this article, we showed that the Mori--Zwanzig formalism with Mori's projector \cite{mori1965transport} is not only consistent with but also a generalization of the existing Koopman learning procedures, such as the extended dynamic mode decomposition (EDMD, \cite{williams2015data}). The propagator of the projected image $\bfvC\l(\Delta\r)\cdot \bfvC^{-1}\l(0\r)$ for any time separation $\Delta$ is identified as the discrete-time approximate Koopman operator $\bfvK^\text{Koop}_\Delta$. We identify that the propagator does not only contain the effect of the instantaneous Markov transition matrix $\bfvM\equiv \dot{\bfvC}\l(0\r)\cdot \bfvC^{-1}\l(0\r)$, but also the memory effect from the trajectory between $t=0$ to $t=\Delta$ (Lemma \ref{lemma:1}). Such a history dependence emerged because of the dynamics is not fully resolved---that is, the dynamics does not evolves invariantly in functional subspace $\Hilg$ spanned by the selected observables. Although the Markov transition matrix $\bfvM$ is the instantaneous Koopman operator, the finite-time propagator is $\exp\l(\Delta \bfvM\r)$ only when the dynamical system is fully resolved. In those partially observed cases, there always emerge a history-dependent term and the orthogonal dynamics from the unresolved degrees of freedom of the dynamics.

Motivated by the data-driven Koopman learning methods, we constructed two numerical algorithms which extract key operators in the Mori--Zwanzig formalism, specifically, with Mori's projection operator. One of our algorithms extracts the continuous-time operators, and the other the discrete-time ones. To the lowest order, the algorithms are operationally identical to the EDMD and they compute the continuous-time Markov matrix $\bfvM$ and the discrete-time approximate Koopman operator $\bfvO^{(0)}_\Delta$. The novelty of our algorithms is that they go beyond the lowest order and proceed with a recursive procedure which uses further two-time correlations to extract the memory kernels (continuous-time $\bfvK(s)$ and discrete-time $\bfvO^{(\ell)}_\Delta$, $\ell\ge 1$). The orthogonal dynamics can be computed after the Markov transition matrix and the memory kernel are obtained. 

Our proposed algorithms provide an alternative data-driven way of applying the Mori--Zwanzig formalism to study dynamical systems. Our approach bypasses the conventional need of modeling the memory kernel and the orthogonal dynamics. The numerical analysis on the test problem shows the complex behavior of the memory kernel which is not likely to be modeled by simple mathematical models. Importantly, we numerically verified that the extracted memory kernel and orthogonal noise satisfy the self-consistent Generalized Fluctuation-Dissipation relationship. To our knowledge, it is the first time that the GFD is numerically verified on a nontrivial model whose analytic solution is not known. We are confident with the validity of the extracted memory kernel because of the rather stringent GFD relationship. 

We showed that the Mori--Zwanzig formalism with the numerically extracted memory kernel and the past history can significantly improve the accuracy of the prediction of the Extended Dynamic Mode Decomposition \cite{williams2015data}. The reason is that the past history contains partial information of the orthogonal dynamics, $\bfvg_\perp(t)$, and the prediction can be improved by incorporating the information. We remark that such a memory-dependent learning is fundamentally different from the recently proposed time-embedded Koopman learning methods \cite{horenko07data,Arbabi17ergodic,brunton2017chaos,clainche17higher,kamb20time,wanner2020robust,Lin2021} which include the past history in the set of the observables. These methods, motivated by the famous Takens' embedding theorem \cite{takens81detecting}, requires to expand the number in the set of variables by $h$ times, where $h$ is the number of the past snapshots and $h$ has to be specified before the computation. With such a construction, the subspace that functions are projected to is $\text{span}(\l\{ \bfvg\l(\l(h-1\r)\Delta\r)\ldots  \bfvg(\Delta), \bfvg(0) \r\})$ which is larger than $\Hilg$. Despite the one-step projection using the past history seems to be mapped by a memory kernel, a more proper way to interpret the operation is to regard the augmented configuration mapped by the Markov transition matrix to that at the next discrete time, i.e. $\l[\bfvg\l(h\Delta\r)\ldots  \bfvg(2 \Delta), \bfvg(\Delta)\r]^T = \bfvM \cdot \l[\bfvg\l(h\Delta\r)\ldots  \bfvg(2 \Delta), \bfvg(\Delta)\r]^T  + \bfvW $ where $\bfvW_0$ is the orthogonal noise. Note that with a finite ($h<\infty$) past history, the Mori--Zwanzig memory kernel would emerge if a history is longer than $h$ steps are given and if the past $h$-step observables do not linearly span an invariant manifold. Time-embedded analysis requires a more expensive inversion of the larger $\bfvC(0) \in \mathbb{R}^{Mh\times Mh} $ matrix. In comparison, the memory kernel of the Mori--Zwanzig construction requires a single inversion of the $\bfvC(0) \in \mathbb{R}^{M\times M}$ and the memory kernel is constructed by other two-time correlations $\bfvC(k\Delta)$. 

Our approach here is close to but different from the method presented by Lin and Lu \cite{Lin2021}, and it merits a more careful and detailed comparison. At the very high level, we share the same formal construction of the discrete-time GLE. Our approaches diverged as soon as the projection operator is chosen: we chose Mori's projection (referred to as the finite-rank projection in \cite{Lin2021}) and Lin an Lu chose the Wiener projection. Formally, choosing Wiener projection is identical to the time-embedding technique: as pointed out in \cite{Lin2021}, the subspace to which Wiener projection projects is spanned by the past trajectory, and there is no Mori--Zwanzig memory kernel for a single prediction.  Lin and Lu had to impose a ``decaying memory condition'', by which they meant the Markov transition weight of distant past configuration must decay. In our case, the Mori--Zwanzig memory kernel naturally decays and no such a constraint is imposed. Computationally, our operations are always linear operations with explicit solution of the linear optimizer, and nonlinear optimization is needed in \cite{Lin2021} because they invoke the rational approximations after the $z$-transformation. In this manuscript, we did not propose practical ways to model the orthogonal dynamics $\bfvW_k$, but Lin and Lu provided a practical way forward using multivariate Gaussian process model to match the power spectrum. As time-embedded analysis are much more expensive to fit but have the benefit of having smaller prediction error [data not shown], it remains an interesting future research direction to objectively evaluate these two methods: under the same computational budget, which one has a higher accuracy and with what metric (e.g. error in prediction, error in estimated Koopman spectrum, etc.)? Which one converge faster with the same finite data set? 

Although the mathematical construction of the Mori--Zwanzig formalism and Koopman theory is general for any set of observables, the accuracy of the prediction crucially depends on the selection of the observables. In practice, it is preferable to adopt a set of observables which invoke smaller  orthogonal dynamics $\bfvF(t)$. Our algorithms, which are first to our knowledge, extract the exact orthogonal dynamics from the data and thus they provide an opportunity for us to study the statistics from the extracted orthogonal dynamics. One possibility in the future is to treat the orthogonal dynamics, $\bfvF(t)$, as another dynamical system and recursively perform Mori--Zwanzig learning to telescope into the residual dynamics. We propose another possibility, analogous to Gram--Schmidt process, to use the numerically extracted $\bfvF(t)$  for identifying those predominant observables orthogonal to the existing set of observables. We expect by including these orthogonal observables, the predictive accuracy of the model can be improved. We remark that it is much cheaper to extract the memory kernel $\bfvK(s)$ than the orthogonal dynamics $\bfvF(t)$. Since the memory kernel is related to the two-time correlation function of the orthogonal dynamics by the Generalized Fluctuation-Dissipation relationship, it will be an interesting research direction to identify those properties of the memory kernel that can be directly used for optimizing the observables, potentially combining the recent development of using deep neural networks as approximate functions by Yeung et al.~\cite{Yeung19}. The extracted operators provides multiple angles for such an optimization: For example, should the objective be minimizing the magnitude of the memory kernel (which $\propto \l\langle \bfvF(t), \bfvF^{T}(0)\r\rangle$ by Eq.~\eqref{eq:GFD}), or the  timescale of the memory kernel as one would like to achieve in the Markov State Models? 

We are currently working on generalizing the proposed methods to partially observed Markov stochastic systems with a focus on spectral analysis on the Mori--Zwanzig operators. In terms of applications, we are applying the proposed data-driven learning algorithm to study the modeling of turbulent flows  \cite{Tian2021datadriven} and molecular dynamical systems, both which are extremely challenging  engineering applications.

\section*{Acknowledgments}
This work has been authored by employees of Triad National Security, LLC which operates Los Alamos National Laboratory (LANL) under Contract No. 89233218CNA000001 with the U.S. Department of Energy/National Nuclear Security Administration. 
The work has been supported by LDRD (Laboratory Directed Research and Development) program at LANL under project 20190059DR (Machine Learning for Turbulence). YTL was partially supported by project LDRD-20190034ER (Massively Parallel Acceleration of the Dynamics of Complex Systems: a Data-Driven Approach) for finalizing the manuscript. YTL sincerely appreciates many insightful discussions with Danny Perez. 

\bibliographystyle{siamplain}

\appendix

\section{Solution of the orthogonal component} \label{app:orthogonal}

It is straightforward to check that \eqref{eq:gperpsolution}, 
\begin{equation}
\bfvg_\perp(t) = \int_0^t \bfvC(t-s) \cdot \bfvC^{-1} \l(0\r)\cdot \bfvF(s) \dd s
\end{equation}
is the solution to \eqref{eq:OE}:
\begin{align}
    \frac{\dd}{\dd t} \bfvg_\perp(t) ={}& \frac{\dd}{\dd t} \l[\int_0^t \bfvC(t-s) \cdot \bfvC^{-1} \l(0\r)\cdot \bfvF(s) \dd s\r]  \\
    ={}& \bfvF(t) + \int_0^t \frac{\dd \bfvC\l(t-s\r)}{\dd t}  \cdot \bfvC^{-1} \l(0\r)\cdot \bfvF(s) \dd s, \nonumber
\end{align}
and by Eq.~\eqref{eq:CE}, 
\begin{align}
    \frac{\dd}{\dd t} \bfvg_\perp(t) ={}& \bfvF(t) + \int_0^t \bfvM\cdot \bfvC\l(t-s\r) \cdot \bfvC^{-1} \l(0\r)\cdot \bfvF(s) \dd s \\
     {}& -  \int_0^t \int_0^{t-s} \bfvK \l( w \r) \cdot \bfvC\l(t-s-w\r) \cdot \bfvC^{-1} \l(0\r)\cdot \bfvF(s) \dd w \dd s\nonumber.
\end{align}
We assume that the integrand satisfies the conditions which allow the change of the order of integrations:
\begin{align}
    \frac{\dd}{\dd t} \bfvg_\perp(t) ={}& \bfvF(t) + \bfvM\cdot \int_0^t  \bfvC\l(t-s\r) \cdot \bfvC^{-1} \l(0\r)\cdot \bfvF(s) \dd s \\
     {}& -  \int_0^t  \bfvK \l( w \r) \l[\int_0^{t-w} \cdot \bfvC\l(t-w-s\r) \cdot \bfvC^{-1} \l(0\r)\cdot \bfvF(s) \dd s \r] \dd w  \nonumber \\
     = {}& \bfvF(t) + \bfvM\cdot\bfvg_\perp\l(t\r) -  \int_0^t  \bfvK \l( w \r) \bfvg_\perp \l(t-w\r) \dd w, \nonumber 
\end{align}
which is Eq.~\eqref{eq:gperpsolution}. In addition, $\bfvg_\perp(t)=0$ satisfies the initial condition in \eqref{eq:OE}. 

\section{Derivation of the discrete-time Mori--Zwanzig formalism from continuous-time} \label{app:discrete-time}

We present two independent derivations of the discrete-time Mori--Zwanzig formalism, assuming the underlying process is a continuous-time and deterministic dynamics. In the first approach presented in \ref{app:discrete-time-1}, we begin with the continuous-time equations \eqref{eq:GLE}, \eqref{eq:CE}, and \eqref{eq:OP}, solving for the quantities evaluated on a temporally evenly sampled time grid $t=0, \Delta, 2\Delta \ldots$, and identify the recursive relationship between the solutions. In the second approach presented in \ref{app:discrete-time-2}, we consider to discretize the dynamics first by transforming the continuous-time operator to a abstract discrete-time map, and re-derive the Mori--Zwanzig equations (analogous to  \eqref{eq:GLE}, \eqref{eq:CE}, and \eqref{eq:OP}) to the discrete-time map. Importantly, in both approaches, we do not impose infinitesimal constraint on $\Delta$, and it can be any finite number. 

Interestingly, the two constructs delivers the same recursive relationship that resembles the continuous-time Mori--Zwanzig equations. We shall refer to these relationships as the discrete-time Mori--Zwanzig equations. However, there exists a subtle difference that the same discrete-time Mori--Zwanzig equations link different mathematical objects in the two approaches. We discuss this subtle difference between the two approaches, and a sufficient condition that the two descriptions agree, in \ref{app:discrete-time-diff}. The generalized fluctuation-dissipation relationship in the discrete-time formulation is discussed in \ref{app:discrete-time-GFD}.

\subsection{Discretization of the continuous-time Mori--Zwanzig equations} \label{app:discrete-time-1}

The Generalized Langevin Equation \eqref{eq:GLE}, the evolutaionary equations for the correlation matrix $\bfvC(t)$ (Eq.~\eqref{eq:CE}) and the projected image $\mathcal{P} \bfvg(t)=\bfvg_{\parallel}(t)$ (Eq.~\eqref{eq:OP}) exhibit a similar evoluationary operator which involves the Markov matrix $\bfvM$ and the memory kernel $\bfvK(t)$. Thus, here we consider the evolutionary operator applied on a test matrix $\bfvT(t)$, which is $M \times  P$ where $M$ is the number of the observables serving as our basis functions spanning $\Hilg$. In the case of Eqs.~\eqref{eq:GLE} and \eqref{eq:CE}, $P=1$, and in the case of Eq.~\eqref{eq:CE}, $P=M$. Suppose the evolutionary equation of the test matrix $\bfvT(t)$ satisfies
\begin{equation}
\frac{\dd}{\dd t} \bfvT(t) =  {\bf M} \cdot \bfvT (t) - \int_0^t \bfvK\l(t-s\r) \cdot \bfvT\l(s\r) \dd s, \label{eq:evoT}
\end{equation}
and suppose we only measure the snapshots of the observables at times on a evenly spaced grid $t=\l[0, \Delta, 2\Delta \ldots\r]$ with a not necessary small separation $\Delta$. We also assume that we know the initial value $\bfvT(0)$. The central aim of this section is to prove that the solution $T(k \Delta )$, $k\in \mathbb{Z}_+$, can be written as
\begin{equation}
\bfvT\l(\l(k+1\r)\Delta\r) = \sum_{\ell=0}^{k} \bfvO^{\l(k\r)} \cdot \bfvT \l(\l(k-\ell\r) \Delta \r), 
\end{equation}
where $\bfvO^{\l(k\r)} $ are prescribed $M\times M$ matrices. We break down the proof into a few separate Lemmas. 

\begin{lemma}\label{lemma:1}
Given the evolutionary equation \eqref{eq:evoT} with an initial condition $\bfvT(0)$ and the continuous-time kernel $\bfvK(s)$, for any $t\ge 0$, the solution $\bfvT(t)$ can be expressed as a linear operator $\bfvO_t$ parametrized by the continuous-time $t$ operated on the initial condition $\bfvT(0)$:
\begin{equation}
\bfvT\l(t\r) = \bfvO_t^{(0)} \cdot \bfvT\l(0\r). 
\end{equation}
\end{lemma}

\begin{proof}
We observe that the correlation matrix $\bfvC(t)$ satisfies Eq.~\eqref{eq:CE}, which has a similar form to Eq.~\eqref{eq:evoT}. It is easy to check that $\bfvO^{0}_t:=\bfvC(t)\cdot\bfvC^{-1}(0)$ is the solution: 
\begin{align}
    \frac{\dd}{\dd t} \bfvT(t) = {}&  \frac{\dd}{\dd t} \l[ \bfvC(t)\cdot\bfvC^{-1}(0) \cdot \bfvT(0) \r] \\
    ={}& \l[ \frac{\dd}{\dd t}\bfvC(t) \r]\cdot \bfvC^{-1}(0) \cdot \bfvT(0) \nonumber \\ 
    ={}& \l[ {\bf M} \cdot \bfvC (t) - \int_0^t \bfvK\l(t-s\r) \cdot \bfvC\l(s\r) \dd s \r]\cdot\bfvC^{-1}(0) \cdot \bfvT(0) \nonumber \\
    ={}& {\bf M} \cdot \bfvT (t) - \int_0^t \bfvK\l(t-s\r) \cdot \bfvT\l(s\r) \dd s. \nonumber
\end{align}

\end{proof}

\begin{theorem} \label{thm:1}
Given (1) the evolutionary equation \eqref{eq:evoT}, (2) the continuous-time memory kernel $\bfvK(s)$, $s\ge 0$, (3) a non-negative integer $k\in \mathbb{Z}_+$, and (4) snapshots of past history at discrete times $\bfvT(j\Delta)$, $ j=0,1,\ldots \le k$, we can express $\bfvT$ at a future time $k\Delta + \tau$, $0 < \tau \le \Delta$ in terms of linear superpositions of the past snapshots:
\begin{equation}
\bfvT\l(\tau + k \Delta\r) = \sum_{\ell=0}^{k} \bfvO_\tau^{\l(\ell\r)} \cdot \bfvT \l(\l(k-\ell\r) \Delta \r), \label{eq:discreteGLEtau}
\end{equation}
where the higher order operators $\bfvO_\tau^{\l(k\r)}$, $k \in \mathbb{N}_+$ are recursively defined by
\begin{equation}
\bfvO_\tau^{\l(k\r)} = \bfvO_{ \tau+k\Delta }^{\l(0\r)} - \sum_{\ell=0}^{k-1} \bfvO_{\tau }^{\l(\ell\r)} \bfvO_{\l(k-\ell \r) \Delta}^{\l(0\r)}. \label{eq:recursiveOperators}
\end{equation}
\end{theorem}

\begin{proof}
From Lemma \ref{lemma:1},
\begin{equation}
\bfvT(\tau + k\Delta) = \Omega_{\tau+k\Delta}^{\l(0\r)} \bfvT(0),
\end{equation}
and the recursive relationship \eqref{eq:recursiveOperators} states
\begin{equation}
\bfvO_{\tau + k \Delta  }^{\l(0\r)} = \sum_{\ell=0}^{k} \bfvO_\tau^{\l(\ell\r)}  \bfvO_{\l(k-\ell \r) \Delta}^{\l(0\r)},
\end{equation}
and thus,
\begin{equation}
\bfvT(\tau + k\Delta) =  \sum_{\ell=0}^{k} \bfvO_\tau^{\l(\ell\r)}  \bfvO_{\l(k-\ell \r) \Delta}^{\l(0\r)}  \bfvT(0) =  \sum_{\ell=0}^{k} \bfvO_\tau^{\l(\ell\r)}  \bfvT(\l(k-\ell\r) \Delta).
\end{equation}
\end{proof}

\begin{corollary}
Given the operators \eqref{eq:recursiveOperators}, $\bfvT\l(\l(k+1\r)\Delta\r)$ can be expressed as linear combination of the past snapshots $\bfvT\l(\ell \Delta \r)$, $\ell=0,1,\ldots k$:
\begin{equation}
\bfvT\l( \l(k+1\r)\Delta \r)  =   \sum_{\ell=0}^{k} \bfvO_\Delta^{\l(\ell\r)}  \bfvT(\l(k-\ell\r) \Delta).
\end{equation}
\end{corollary}

\begin{corollary}
Because the correlation matrix $\bfvC(t)$ satisfies Eq.~\eqref{eq:CE} which is of the form \eqref{eq:evoT}, given the operators \eqref{eq:recursiveOperators}, the snapshots of the correlation matrix at discrete times satisfy
\begin{equation}
\bfvC\l( \l(k+1\r)\Delta \r)  =   \sum_{\ell=0}^{k} \bfvO_\Delta^{\l(\ell\r)}  \bfvC(\l(k-\ell\r) \Delta). \label{eq:discrete-CE-1}
\end{equation}
\end{corollary}

\begin{corollary}\label{cor:1}
Because the projected image  $\mathcal{P}\bfvg(t)$ satisfies Eq.~\eqref{eq:OP} which is of the form \eqref{eq:evoT}, given the operators \eqref{eq:recursiveOperators}, the snapshots of the projected image at discrete times satisfy
\begin{equation}
\mathcal{P}\bfvg\l( \l(k+1\r)\Delta \r)  =   \sum_{\ell=0}^{k} \bfvO_\Delta^{\l(\ell\r)}  \mathcal{P}\bfvg(\l(k-\ell\r) \Delta).  \label{eq:discrete-OP-1}
\end{equation}
\end{corollary}

\begin{theorem}
Discretized Generalized Langiven Equation. Given the GLE \eqref{eq:GLE} and the associated operators \eqref{eq:recursiveOperators}, the snapshots $\bfvg(t)$ at discrete times $(k+1)\Delta$, $k\in \mathbb{N}$, satisfy
\begin{equation}
\bfvg\l( \l(k+1\r)\Delta \r)  =   \sum_{\ell=0}^{k} \bfvO_\Delta^{\l(\ell\r)}  \bfvg(\l(k-\ell\r) \Delta) + \bfvW_{k}.  \label{eq:discrete-GLE-1}
\end{equation}
where $\bfvW_{k}$ is the discrete-time orthogonal dynamics, which is a linear function of the orthogonal dynamics $\bfvF(t)$, $t\le (k+1)\Delta$ and $\bfvW_{k+1}$ is orthogonal to $\Hilg$. 
\end{theorem}
\begin{proof}
As illustrated in Sec.~\ref{sec:geometry}, the solution of GLE can be written as a general solution $\bfvg_\parallel(t)$ and a particular solution $\bfvg_\perp(t)$ satisfying Eqs.~\eqref{eq:PE} and \eqref{eq:OE} respectively. Note that $\bfvg_\parallel(t)$ is just the projected image $\mathcal{P}\bfvg(t)$, and from Corollary \ref{cor:1}, we conclude that
\begin{equation}
\bfvg_\parallel\l(\l(k+1\r) \Delta\r) = \sum_{\ell=0}^{k} \bfvO^{\l(\ell\r)}_\Delta\cdot \bfvg_\parallel \l(\l(k-\ell\r)\Delta \r).
\end{equation}
Because $\bfvg(t) = \bfvg_\parallel(t) + \bfvg_\perp(t) $, $\forall t\ge 0$, at time $t=\l(k+1\r) \Delta$, we can equate
\begin{align}
\bfvg\l(\l(k+1\r)\Delta \r) ={}& \bfvg_\parallel\l(\l(k+1\r)\Delta \r) + \bfvg_\perp \l(\l(k+1\r)\Delta \r)  \\
={}& \sum_{\ell=0}^{k} \Omega^{\l(\ell\r)}_\Delta\cdot \bfvg_\parallel \l(\l(k-\ell\r)\Delta \r)  + \bfvg_\perp \l(\l(k+1\r) \Delta \r) \nonumber \\
={}& \sum_{\ell=0}^{k} \Omega^{\l(\ell\r)}_\Delta\cdot \l[ \bfvg_\parallel \l(\l(k-\ell\r)\Delta \r) +  \bfvg_\perp \l(\l(k-\ell\r)\Delta \r)\r]  + \bfvg_\perp \l(\l(k+1\r) \Delta \r)  \nonumber \\{}&-  \sum_{\ell=0}^{k} \Omega^{\l(\ell\r)}_\Delta\cdot  \bfvg_\perp \l(\l(k-\ell\r)\Delta \r) \nonumber \\
={}& \sum_{\ell=0}^{k} \Omega^{\l(\ell\r)}_\Delta\cdot \bfvg\l(\l(k-\ell\r)\Delta \r) \nonumber \\{}& +  \l[\bfvg_\perp \l(\l(k+1\r) \Delta \r)  -  \sum_{\ell=0}^{k} \Omega^{\l(\ell\r)}_\Delta\cdot  \bfvg_\perp \l(\l(k-\ell\r)\Delta \r) \r].\nonumber
\end{align}
We identify
\begin{equation}    
\bfvW_{k} = \bfvg_\perp \l(\l(k+1\r) \Delta \r) - \sum_{\ell=0}^k \Omega^{\l(k-\ell\r)}_\Delta \bfvg_\perp\l(\ell \Delta\r). \label{eq:discreteNoiseW}
\end{equation}
Note that $\bfvW_{k}$ is a linear function of the snapshots of $\bfvg_\perp$, which are linear functions of $\bfvF(s)$, $t\le \l(k+1\r)\Delta$, and thus $\bfvW_{k}$ is orthogonal to $\Hilg$. 
\end{proof}

\subsection{Mori--Zwanzig equations of the generic discrete-time dynamics} \label{app:discrete-time-2}
We present an intuitive derivation of that is analogous to Sec.~\ref{sec:intuitiveDerivation}, noting that a more general derivation can be found in reference \cite{Lin2021}. We begin with Eq.~\eqref{eq:fullKoopman}, and integrate the equation to the discrete time grid $t=0, \Delta, 2 \Delta, \ldots$ to obtain the discrete mapping in the full Hilbert space $\Hil$:
\begin{equation} 
\begin{bmatrix}
\bfvg_\MM(t+\Delta) \\
\bfvg_\MMbar(t+\Delta) 
\end{bmatrix}
=e^{\Delta \bfvL }\cdot \begin{bmatrix}
\bfvg_\MM(t) \\
\bfvg_\MMbar(t) 
\end{bmatrix}
:=
 \begin{bmatrix}
\bfv{U}_{\MM\MM} &  \bfv{U}_{\MM\MMbar} \\
\bfv{U}_{\MMbar\MM} & \bfv{U}_{\MMbar\MMbar}
\end{bmatrix}
\cdot
 \begin{bmatrix}
\bfvg_\MM(t) \\
\bfvg_\MMbar(t) 
\end{bmatrix}. 
\label{eq:intuitiveDiscreteMapping}
\end{equation}
Given the initial condition $\bfvg_\MM\l(0\r)$ and $\bfvg_\MMbar\l(0\r)$, the solution of the orthogonal component at the discrete times, $\bfvg_\MMbar\l(k \Delta\r)$, can be expressed in terms of the historical snapshots of the resolved components, $\bfvg_\MM\l(\ell \Delta\r)$, $\ell = 0, 1, \ldots k$, and the initial condition $\bfvg_\MMbar\l(0\r)$:
\begin{equation}
\bfvg_\MMbar\l( \l(k+1\r) \Delta\r) = \sum_{\ell=0}^{k}  \bfv{U}_{\MMbar\MMbar}^{\ell}  \bfv{U}_{\MMbar\MM} \bfvg_\MM\l(\l(k-\ell\r)\Delta \r) +  \bfv{U}_{\MMbar\MMbar}^{k} \bfvg_\MMbar\l(0\r),
\end{equation}
thus,
\begin{align}
\bfvg_\MM( \l(k+1\r) \Delta)  = {}&  \bfv{U}_{\MM\MM}  \bfvg_\MM( k \Delta) +  \bfv{U}_{\MM\MMbar}  \bfvg_\MMbar( k \Delta)  \\
={}&  \bfv{U}_{\MM\MM}  \bfvg_\MM( k \Delta) +   \bfv{U}_{\MM\MMbar}  \bfv{U}_{\MMbar\MMbar}^{k} \bfvg_\MMbar\l(0\r) \nonumber \\
{}&+  \bfv{U}_{\MM\MMbar}   \sum_{\ell=0}^{k-1}  \bfv{U}_{\MMbar\MMbar}^{\ell}  \bfv{U}_{\MMbar\MM} \bfvg_\MM\l(\l(k-\ell-1\r)\Delta \r).\nonumber
\end{align}
Now we define $\bfv{\Lambda}_\Delta^{\l(0\r)}:= \bfv{U}_{\MM\MM} $,  $\bfv{\Lambda}_\Delta^{\l(\ell\r)}:= \bfv{U}_{\MM\MMbar}  \bfv{U}_{\MMbar\MMbar}^{(\ell-1)}  \bfv{U}_{\MMbar\MM} $, $\ell=1,2,\ldots$, and $\bfv{V}_k :=   \bfv{U}_{\MM\MMbar}  \bfv{U}_{\MMbar\MMbar}^{k} \bfvg_\MMbar\l(0\r)$ and obtain the discrete-time generalized Langevin equation 
\begin{equation}
\bfvg\l( \l(k+1\r)\Delta \r)  =   \sum_{\ell=0}^{k} \bfv{\Lambda}_\Delta^{\l(\ell\r)}  \bfvg(\l(k-\ell\r) \Delta) + \bfv{V}_{k}. \label{eq:discrete-GLE-2}
\end{equation}

A more general derivation first transforms the dynamics Eq.~\eqref{eq:phiEvo} to a discrete map of the solutions evaluated at $t=0, \Delta, 2\Delta \ldots$,
\begin{equation}
\bfvPhi(t+\Delta) = \bfv{U}_\Delta \l( \bfvPhi(t) \r)
\end{equation}
Here, $\bfv{U}_\Delta$ is the nonlinear operator defined as the solution of the continuous-time equation:
\begin{equation}
\bfv{U}_\Delta \l( \bfvPhi_0 \r) := \int_0^\Delta \bfvR\l(\bfvPhi\l(t\r)\r) \dd t + \bfvPhi_0. 
\end{equation}
With the transformed discrete-time map, $\bfv{U}_\Delta$, we apply the generic Mori--Zwanzig formulation for the discrete-time dynamics \cite{Lin2021} to obtain the discrete-time GLE \eqref{eq:discrete-GLE-2}. 

Because the samples collected from this picture involves only discrete-time snapshots separated by a finite time $\Delta$, we need to replace the inner product from averaging over a continuous-time trajectory (Eq.~\eqref{eq:ctsInnerProduct}) by a discrete-time sum:
\begin{equation}
\l\langle f , h\r\rangle_\Delta := \lim_{N\rightarrow \infty }\frac{1}{N} \sum_{i=1}^{N} \int_{\Omega} f\l(\bfvPhi_0\r) h\l(\bfvPhi_0\r) \dd \mu\l(\bfvPhi_0\r). \label{eq:disInnerProduct}
\end{equation}
We put a subscript under the inner product $\l\langle \cdot , \cdot \r\rangle_\Delta$ to denote the difference between \eqref{eq:disInnerProduct} and its continuous-time counterpart \eqref{eq:ctsInnerProduct}. 

Finally, a similar analysis to the one we presented in Sec.~\ref{sec:CE+OP} results in the discrete-time recursive relationship between the correlation matrix $\bfvC_\Delta \l( k \Delta \r) $ and the projected image $\mathcal{P}_\Delta \bfvg\l(k \Delta \r) $, $k=0,1,2\ldots$,
\begin{align}
\bfvC_\Delta \l( \l(k+1\r)\Delta \r)  ={}&   \sum_{\ell=0}^{k} \bfv{\Lambda}_\Delta^{\l(\ell\r)}  \bfvC_\Delta (\l(k-\ell\r) \Delta), \label{eq:discrete-CE-2} \\
\mathcal{P}_\Delta \bfvg\l( \l(k+1\r) \Delta \r)    ={}&   \sum_{\ell=0}^{k} \bfv{\Lambda}_\Delta^{\l(\ell\r)}  \mathcal{P}_\Delta \bfvg\l( \l(k-\ell\r) \Delta \r). \label{eq:discrete-OP-2}
\end{align}
Again, we put the subscript to the correlation matrix $\bfvC_\Delta$ and the projection operator $\mathcal{P}_\Delta$ to differentiate them from their counterparts computed with continuous-time inner product, Eq.~\eqref{eq:ctsInnerProduct}.

\subsection{Difference between the two formulations}\label{app:discrete-time-diff}

The discrete-time dynamics in the above two formulations coincide, and it is tempting to equate the operators $\bfvO_\Delta^{\l(\ell\r)}$  to $\bfv{\Lambda}_\Delta^{\l(\ell\r)}$, and the discrete-time noise $\bfvW_k$ to $\bfv{V}_k$. Nevertheless, there is a subtle difference between these two formulation, and in general, they do not have to be the same. The subtlety is that the inner product is defined by integrating over a continuous domain of time in the first formulation, but by summing over a discrete domain of the time in the second derivation. The invariant measure of the former does not have to be equal to the latter. Consequently, the projection operator, which depends on the definition inner product, Eq.~\eqref{eq:innerProduct}, does not have to be identical in these two formulation. 

In our proposed algorithm \ref{alg:2}, the discrete-time operators (either $\bfvO_\Delta^{\l(\ell\r)}$ or $\bfv{\Lambda}_\Delta^{\l(\ell\r)}$) and noise (either $\bfvW_k$ or $\bfv{V}_k$) are extracted from the correlation matrices. In the first formulation, the correlation matrix was computed with the inner product Eq.~\eqref{eq:ctsInnerProduct},
\begin{equation}
\bfvC\l(k\Delta \r) := \lim_{T\rightarrow \infty} \frac{1}{T} \int_0^{k\Delta } \bfvg \circ \bfvPhi\l(k\Delta+s\r) \cdot \bfvg^T \circ \bfvPhi\l(s\r) \dd s, \label{eq:ctsC} 
\end{equation}
and in the second approach, it is computed by with the inner product Eq.~\eqref{eq:disInnerProduct},
\begin{align}
\bfvC_\Delta\l(k\Delta \r) :={}&  \lim_{N\rightarrow \infty}\frac{1}{N} \sum_{i=0}^{N-1} \bfvg \circ \bfvPhi\l(\l(k+i\r) \Delta \r) \cdot \bfvg^T \circ \bfvPhi\l( i\Delta \r) . \label{eq:disC} 
\end{align}

Our analysis shows that a sufficient condition for $\bfvO_\Delta^{\l(\ell\r)}=\bfv{\Lambda}_\Delta^{\l(\ell\r)}$ and $\bfvW_k=\bfv{V}_k$ is $\bfvC\l(k\Delta \r)=\bfvC_\Delta\l(k\Delta \r)$. The correlation matrices computed by two approaches are not necessarily identical. A simple harmonic oscillator $\dot{x}=p$ and $\dot{p}=-x$ with a unit amplitude $x^2(t)+p^2(t)=1$ can be a counterexample. It has a period $2\pi$, and if we choose $\Delta$ is $2\pi/3$, these two formulations can be quite different: the continuous  correlation matrix (Eqs.~\eqref{eq:ctsC}) has $\l\langle x, x \r\rangle = \pi$, but the discrete  correlation matrix (\eqref{eq:disC}) has $\l\langle x, x \r\rangle = 1 + \cos^2\l(2\pi/3\r) + \cos^2\l(4\pi/3\r) = 3/2$ if $x(0)=1$. Note that in prediction (cf.~Sec.~\ref{sec:errorAnalysis}), the correlation matrix obtained from the first approach can be used to project any configuration satisfying $x^2(0)+p^2(0)=1$, but the correlation matrix obtained from the second approach can be only used to a subset of possible initial conditions: $x(0)\in\l\{0,2\pi/3,4\pi/3\r\}$.

The subtlety between the two formulations serves as a caution to the practitioners to pay attention to formulating \emph{what we aim to learn}. On the one hand, the first formulation presented in \ref{app:discrete-time-1} learns the discrete-time operators $\bfvO_\Delta^{(\ell)}$, which have direct connection to the continuous-time operators because they are defined in terms of the continuous-time Markov transition $\bfvM$ and the memory kernel $\bfvK(t)$. Despite the fact that the correlation matrix is discretized and stored with a coarser resolution, the overall learned operators $\bfvO_\Delta^{(\ell)}$ serve a similar role to the continuous-time Mori--Zwanzig operators. For example, the operators predict the same $\mathcal{P}\bfvg(t)$ in $\Hilg$. Thus, if one aims to learn about the continuous-time operators, the first approach is more desirable. The cost of the first approach is that even though the discretization $\Delta$ can be finite, one still needs a very finely sampled temporal grid to evaluate the continuous-time correlation matrix $\bfvC(t)$, if an analytical expression of $\bfvC(t)$ is not possible. We remark that an additional online computation of $\bfvC$ in the fine-scale simulation can make computations efficient. On the other hand, a discretization with a finite $\Delta$ is beneficial because it provides another coarser resolution in which the correlation matrix is stored.  The second formulation presented in \ref{app:discrete-time-2} learns the Mori--Zwanzig operators of the \emph{generic discrete maps}. In those cases where the inner product defined by the discrete-time statistics (\eqref{eq:disC}) is not identical to the continuous-time statistics (\eqref{eq:ctsC}), the projection operator in the second formulation is not the same as the one in the continuous-formulation.  Consequently, the operators would predict a different projected image $\mathcal{P}_\Delta\bfvg(t)$ in $\Hilg$. The above simple harmonic oscillator provides an intuitive example.

Finally, we point out that despite the subtle differences, if the two measures (statistics from discrete-time snapshots and continuous-time dynamics) are identical, these two formulations converge. In this scenario, there is no difference between the formulations in the theoretical sense. Nevertheless, computationally, the first approach---choosing a small $\delta$ to collect the snapshot data, computing the correlation matrix $\bfvC$, and then discretizing $\bfvC$ to multiples of $\delta$---possesses two advantages. First, because the computation of the correlation matrix is agnostic to the discretization parameter $\Delta$, one does not need to re-simulate the discrete-time simulation when we change $\Delta$. Secondly, recall that we need to compute the correlation matrix from the snapshots, which were recorded from a high-fidelity simulation which is generally computationally expensive. Thus, given a fixed amount of computational resource, there is generally an upper bound of the physical time which we are allowed to simulate. Because the first approach uses an equal or finer temporal resolution in comparison to the second ($\delta \le \Delta$), we will collect more snapshots before the high-fidelity simulation ends. Thus, computationally, the convergence of the correlation matrix is generally better due to more samples. 

\subsection{Generalized Fluctuation--Dissipation Relationship in the discrete-time formulations}\label{app:discrete-time-GFD}
In each of the frameworks presented in Appendices \ref{app:discrete-time-1} and \ref{app:discrete-time-2}, with respect to the corresponding inner product (see discussion in \ref{app:discrete-time-diff}), there exists a Generalized Fluctuation-Dissipation relationship between the discrete-time noise ($\bfvW_k$ or $\bfvV_k$) and the discrete time operators ($\bfvO_\Delta^{(\ell)}$ or $\bfvLamb_\Delta^{(\ell)}$):
\begin{align}
\bfvO^{\l(k\r)}_\Delta={}& - \l\langle \bfvW_{k}, \bfvW_0^T\r\rangle \cdot \bfvC^{-1}\l(-\Delta \r), \quad k\in \mathbb{N}, \label{eq:discrete-time-GFD1} \\
\bfv{\Lambda}^{\l(k\r)}_\Delta={}& - \l\langle \bfv{V}_{k}, \bfv{V}_0^T\r\rangle_\Delta \cdot \bfvC_\Delta^{-1}\l(-\Delta\r), \quad k\in \mathbb{N}. \label{eq:discrete-time-GFD2}
\end{align}
Here, the first inner product in Eq.~\eqref{eq:discrete-time-GFD1} is an integration against the continuous-time trajectory (e.g., Eq.~\eqref{eq:ctsInnerProduct}), and the second inner product in Eq.~\eqref{eq:discrete-time-GFD2} is a discrete sum over the snapshots (e.g., Eq.~\eqref{eq:disInnerProduct}). Note that $\bfvC(-\Delta) \equiv \bfvC^T(\Delta)$ and $\bfvC_\Delta\l(-\Delta\r) = \bfvC_\Delta^T\l(\Delta\r)$ by definitions \eqref{eq:ctsC} and \eqref{eq:disC}. 
In this section, we present the proof to the generalized fluctuation-dissipation relationship, Eqs.~\eqref{eq:discrete-time-GFD1} and \eqref{eq:discrete-time-GFD2}. 

\subsubsection{Discretization of the continuous-time Mori--Zwanzig framework}
From Eq.~\eqref{eq:discreteNoiseW} and noting that $\bfvg_\perp\l(0\r)=0$ (because $\bfvg(0) = \bfvg_\parallel(0)$), 
\begin{align}
\l\langle \bfvW_{k}, \bfvW_0^T\r\rangle  ={}& \l\langle\bfvg_\perp \l(\l(k+1\r) \Delta \r) - \sum_{\ell=0}^k \Omega^{\l(k-\ell\r)}_\Delta \bfvg_\perp\l(\ell \Delta\r), \bfvg^T_\perp(\Delta ) \r\rangle  \\
={}&  \l\langle \bfvg \l(\l(k+1\r) \Delta \r) - \bfvg_\parallel \l(\l(k+1\r) \Delta \r), \bfvg^T(\Delta )-\bfvg^T_\parallel\l( \Delta\r) \r\rangle \nonumber\\{}& - \l\langle \sum_{\ell=0}^k \Omega^{\l(k-\ell\r)}_\Delta \l[\bfvg\l(\ell \Delta\r) - \bfvg_\parallel\l(\ell \Delta\r)\r], \bfvg^T(\Delta )-\bfvg^T_\parallel\l( \Delta\r) \r\rangle. \nonumber
\end{align}
This is because $\bfvg_\perp(t) = \bfvg(t)-\bfvg_\parallel$. Then, we can express $\l\langle \bfvW_{k}, \bfvW_0^T\r\rangle $ in terms of pairs of observables,
\begin{align}
\l\langle \bfvW_{k}, \bfvW_0^T\r\rangle  = {}&
\l\langle \bfvg \l(\l(k+1\r) \Delta \r) , \bfvg^T(\Delta )\r\rangle 
-\l\langle \bfvg \l(\l(k+1\r) \Delta \r) , \bfvg^T_\parallel\l( \Delta\r)  \r\rangle \nonumber \\
{}& -\l\langle \bfvg_\parallel \l(\l(k+1\r) \Delta \r)  , \bfvg^T(\Delta ) \r\rangle 
+\l\langle \bfvg_\parallel \l(\l(k+1\r) \Delta \r)  , \bfvg^T_\parallel\l( \Delta\r) \r\rangle \nonumber \\
{}&-\sum_{\ell=0}^k \Omega^{\l(k-\ell\r)}_\Delta  \l\langle \bfvg\l(\ell \Delta\r) ,  \bfvg^T(\Delta )\r\rangle 
+\sum_{\ell=0}^k \Omega^{\l(k-\ell\r)}_\Delta  \l\langle \bfvg\l(\ell \Delta\r) , \bfvg^T_\parallel\l( \Delta\r)  \r\rangle \nonumber \\
{}&+\sum_{\ell=0}^k \Omega^{\l(k-\ell\r)}_\Delta  \l\langle \bfvg_\parallel\l(\ell \Delta\r), \bfvg^T(\Delta ) \r\rangle 
- \sum_{\ell=0}^k \Omega^{\l(k-\ell\r)}_\Delta  \l\langle \bfvg_\parallel\l(\ell \Delta\r),  \bfvg^T_\parallel\l(\Delta\r) \r\rangle.
\end{align}
The above expression can be simplified by 
\begin{equation}
    \l\langle \bfvg(t), \bfvg^T_\parallel(s)\r\rangle=\l\langle \bfvg_\parallel(t), \bfvg^T_\parallel(s)\r\rangle = \l\langle \bfvg_\parallel(t), \bfvg^T(s)\r\rangle,
\end{equation}
which leads to
\begin{align}
\l\langle \bfvW_{k}, \bfvW_0^T\r\rangle  = {}&
\l\langle \bfvg \l(\l(k+1\r) \Delta \r) , \bfvg^T(\Delta )\r\rangle 
-\l\langle \bfvg_\parallel \l(\l(k+1\r) \Delta \r)  , \bfvg^T_\parallel\l( \Delta\r) \r\rangle \nonumber \\
{}&-\sum_{\ell=0}^k \Omega^{\l(k-\ell\r)}_\Delta  \l\langle \bfvg\l(\ell \Delta\r) ,  \bfvg^T(\Delta )\r\rangle 
+ \sum_{\ell=0}^k \Omega^{\l(k-\ell\r)}_\Delta  \l\langle \bfvg_\parallel\l(\ell \Delta\r),  \bfvg^T_\parallel\l(\Delta\r) \r\rangle \nonumber \\
= {}& \l\langle \bfvg \l(\l(k+1\r) \Delta \r) , \bfvg^T(\Delta )\r\rangle -\sum_{\ell=0}^k \Omega^{\l(k-\ell\r)}_\Delta  \l\langle \bfvg\l(\ell \Delta\r) ,  \bfvg^T(\Delta )\r\rangle,
\end{align}
The last equality comes from Eq.~\eqref{eq:discrete-OP-1}, that 
\begin{align}
 \bfvg_\parallel \l(\l(k+1\r) \Delta \r) -  \sum_{\ell=0}^k \Omega^{\l(k-\ell\r)}_\Delta  \bfvg_\parallel\l(\ell \Delta\r) = 0 \nonumber \\ \Rightarrow
 \l\langle \bfvg_\parallel \l(\l(k+1\r) \Delta \r) -  \sum_{\ell=0}^k \Omega^{\l(k-\ell\r)}_\Delta  \bfvg_\parallel\l(\ell \Delta\r), \bfvg_\parallel^T\l(\Delta \r) \r\rangle = 0.
\end{align}
By definition, $\l\langle \bfvg \l(\l(k+1\r) \Delta \r) , \bfvg^T(\Delta )\r\rangle =\bfvC( k\Delta )$ and $\l\langle \bfvg\l(\ell \Delta\r) ,  \bfvg^T(\Delta )\r\rangle = \bfvC\l(\l(\ell-1\r)\Delta \r)$. Using Eq.~\eqref{eq:discrete-CE-1}, we can further simplify the two-time correlation of the discrete-time noise $\bfvW$:
\begin{align}
\l\langle \bfvW_{k}, \bfvW_0^T\r\rangle  ={}& \bfvC\l( k\Delta \r) - \sum_{\ell=0}^k \Omega^{\l(k-\ell\r)}_\Delta \bfvC\l( \l(\ell-1\r)\Delta \r) \nonumber \\
={}& \sum_{\ell=0}^{k-1}\Omega^{\l(k-1-\ell\r)}_\Delta \bfvC\l( \ell \Delta \r)  - \sum_{\ell=0}^k \Omega^{\l(k-\ell\r)}_\Delta \bfvC\l( \l(\ell-1\r)\Delta \r) \nonumber \\
={}&  \sum_{\ell=0}^{k-1}\Omega^{\l(k-1-\ell\r)}_\Delta \bfvC\l( \ell \Delta \r)  - \sum_{\ell'=-1}^{k-1} \Omega^{\l(k-1-\ell'\r)}_\Delta \bfvC\l( \ell'\Delta \r) \nonumber\\
={}& - \Omega_\Delta^{(k)} \bfvC(\Delta),
\end{align}
and establish Eq.~\eqref{eq:discrete-time-GFD1} by multiplying $\bfvC^{-1}\l(-\Delta\r)$ to both sides of the above equation. 

\subsubsection{Generalized Fluctuation--Dissipation Relationship in the time-discretized dynamics}\label{app:discrete-time-GFD-2}
A parallel analysis to the one presented in the above section \ref{app:discrete-time-GFD} can be carried out to prove Eq.~\eqref{eq:discrete-time-GFD2} with the inner product (e.g., Eq.~\eqref{eq:disC}), the Generalized Langevin Equation \eqref{eq:discrete-GLE-2}, and the evolutionary equations of the correlation matrix Eq.~\eqref{eq:discrete-CE-2} and projected image Eq.~\eqref{eq:discrete-OP-2}.  

An alternative derivation, parallel to the derivation in Sec.~\ref{sec:GFD}, can be carried out. We use the intuitive notation presented in \ref{app:discrete-time-2}. By definition, the discrete-time noise after $k$ snapshots, starting at the $i^\text{th}$ snapshot along the long trajectory is $\bfv{V}_{k\vert i} :=   \bfv{U}_{\MM\MMbar}  \bfv{U}_{\MMbar\MMbar}^{k} \bfvg_\MMbar\l(i\Delta\r)$. Then, the two-time correlation between the noise with respect to the inner product is
\begin{equation}
\l\langle \bfvV_k, \bfvV_0^T \r\rangle = \lim_{N\rightarrow \infty} \frac{1}{N} \sum_{i=1}^N  \bfv{U}_{\MM\MMbar}  \bfv{U}_{\MMbar\MMbar}^{k} \bfvg_\MMbar\l(i\Delta\r) \bfvg^T_\MMbar\l(i\Delta\r) \bfv{U}^T_{\MM\MMbar}.
\end{equation}
From the discrete-time mapping Eq.~\eqref{eq:intuitiveDiscreteMapping}, 
\begin{equation}
    \bfv{U}_{\MM\MMbar} \bfvg_\MMbar\l(i\Delta\r)  = \bfvg_\MM\l(\l(i+1\r) \Delta \r) - \bfv{U}_{\MM\MM} \bfvg_\MM\l(i\Delta\r),
\end{equation} 
and thus,
\begin{align}
\l\langle \bfvV_k, \bfvV_0^T \r\rangle ={}&  \lim_{N\rightarrow \infty} \frac{1}{N} \sum_{i=1}^N  \bfv{U}_{\MM\MMbar}  \bfv{U}_{\MMbar\MMbar}^{k} \bfvg_\MMbar\l(i\Delta\r) \l[\bfvg^T_\MM\l(\l(i+1\r) \Delta\r) - \bfvg^T_\MM\l(i \Delta\r)  \bfv{U}^T_{\MM\MM}\r] \nonumber \\
={}& \lim_{N\rightarrow \infty} \frac{1}{N} \sum_{i=1}^N  \bfv{U}_{\MM\MMbar}  \bfv{U}_{\MMbar\MMbar}^{k} \bfvg_\MMbar\l(i\Delta\r) \bfvg^T_\MM\l(\l(i+1\r) \Delta\r) \nonumber \\
{}& -  \bfv{U}_{\MM\MMbar}  \bfv{U}_{\MMbar\MMbar}^{k} \l\langle \bfvg_\MM, \bfvg^T_\MMbar \r\rangle  \bfv{U}^T_{\MM\MM} \nonumber \\
={}& \lim_{N\rightarrow \infty} \frac{1}{N} \sum_{i=1}^N  \bfv{U}_{\MM\MMbar}  \bfv{U}_{\MMbar\MMbar}^{k} \bfvg_\MMbar\l(i\Delta\r) \bfvg^T_\MM\l(\l(i+1\r) \Delta\r),
\end{align}
because by construction, $\l\langle \bfvg_\MM, \bfvg^T_\MMbar \r\rangle = 0$. Again, using Eq.~\eqref{eq:intuitiveDiscreteMapping}, $\bfv{U}_{\MMbar\MMbar} \bfvg_\MMbar\l(i\Delta\r)  = \bfvg_\MMbar\l(\l(i+1\r) \Delta \r) - \bfv{U}_{\MM\MM} \bfvg_\MM\l(i\Delta\r) $,
\begin{align}
\l\langle \bfvV_k, \bfvV_0^T \r\rangle ={}&  \lim_{N\rightarrow \infty} \frac{1}{N} \sum_{i=1}^N  \bfv{U}_{\MM\MMbar}  \bfv{U}_{\MMbar\MMbar}^{k-1} \bfvg_\MMbar\l(\l(i+1\r) \Delta \r)  \bfvg^T_\MM\l(\l(i+1\r) \Delta\r) \nonumber \\
{}&- \lim_{N\rightarrow \infty} \frac{1}{N} \sum_{i=1}^N  \bfv{U}_{\MM\MMbar}  \bfv{U}_{\MMbar\MMbar}^{k-1} \bfv{U}_{\MM\MM} \bfvg_\MM\l(i\Delta\r) \bfvg^T_\MM\l(\l(i+1\r) \Delta\r) \nonumber \\
={}& \bfv{U}_{\MM\MMbar}  \bfv{U}_{\MMbar\MMbar}^{k-1} \l\langle \bfvg_\MMbar, \bfvg^T_\MM\r\rangle \nonumber \\
{}& - \lim_{N\rightarrow \infty} \frac{1}{N} \sum_{i=1}^N  \bfv{U}_{\MM\MMbar}  \bfv{U}_{\MMbar\MMbar}^{k-1} \bfv{U}_{\MM\MM}  \bfvg_\MM\l(i\Delta\r) \bfvg^T_\MM\l(\l(i+1\r) \Delta\r) \nonumber \\
={}&  -  \bfv{U}_{\MM\MMbar}  \bfv{U}_{\MMbar\MMbar}^{k-1} \bfv{U}_{\MM\MM}  \bfvC_\Delta\l(-\Delta \r).
\end{align}
By definition, $  \bfv{U}_{\MM\MMbar}  \bfv{U}_{\MMbar\MMbar}^{k-1} \bfv{U}_{\MM\MM} = \bfvLamb_\Delta^{\l(k\r)}$,  and thus, we prove Eq.~\eqref{eq:discrete-time-GFD2}:
\begin{equation}
\l\langle \bfvV_k, \bfvV_0^T \r\rangle  = - \bfvLamb_\Delta^{\l(k\r)}  \bfvC_\Delta\l(-\Delta \r) \Rightarrow \bfvLamb_\Delta^{\l(k\r)} =\l\langle \bfvV_k, \bfvV_0^T \r\rangle \bfvC^{-1}_\Delta\l(-\Delta \r). 
\end{equation}

\end{document}